\PassOptionsToPackage{unicode}{hyperref}
\PassOptionsToPackage{hyphens}{url}
\PassOptionsToPackage{dvipsnames,svgnames*,x11names*}{xcolor}
\documentclass[
  american,
]{article}
\usepackage{lmodern}
\usepackage{amssymb,amsmath}
\usepackage{ifxetex,ifluatex}
\ifnum 0\ifxetex 1\fi\ifluatex 1\fi=0 
  \usepackage[T1]{fontenc}
  \usepackage[utf8]{inputenc}
  \usepackage{textcomp} 
\else 
  \usepackage{unicode-math}
  \defaultfontfeatures{Scale=MatchLowercase}
  \defaultfontfeatures[\rmfamily]{Ligatures=TeX,Scale=1}
\fi
\IfFileExists{upquote.sty}{\usepackage{upquote}}{}
\IfFileExists{microtype.sty}{
  \usepackage[]{microtype}
  \UseMicrotypeSet[protrusion]{basicmath} 
}{}
\makeatletter
\@ifundefined{KOMAClassName}{
  \IfFileExists{parskip.sty}{%
    \usepackage{parskip}
  }{
    \setlength{\parindent}{0pt}
    \setlength{\parskip}{6pt plus 2pt minus 1pt}}
}{
  \KOMAoptions{parskip=half}}
\makeatother
\usepackage{xcolor}
\IfFileExists{xurl.sty}{\usepackage{xurl}}{} 
\IfFileExists{bookmark.sty}{\usepackage{bookmark}}{\usepackage{hyperref}}
\hypersetup{
  pdftitle={Weighted quantile estimators},
  pdfauthor={Andrey Akinshin},
  pdflang={en-US},
  colorlinks=true,
  linkcolor=linkcolor,
  filecolor=Maroon,
  citecolor=citecolor,
  urlcolor=urlcolor,
  pdfcreator={LaTeX via pandoc}}
\usepackage[margin=1in]{geometry}
\usepackage{color}
\usepackage{fancyvrb}

\DefineVerbatimEnvironment{Highlighting}{Verbatim}{commandchars=\\\{\}}
\usepackage{framed}
\definecolor{shadecolor}{RGB}{248,248,248}
\newenvironment{Shaded}{\begin{snugshade}}{\end{snugshade}}

\newcommand{\CommentTok}[1]{\textcolor[rgb]{0.56,0.35,0.01}{\textit{#1}}}

\newcommand{\ControlFlowTok}[1]{\textcolor[rgb]{0.13,0.29,0.53}{\textbf{#1}}}
\newcommand{\DataTypeTok}[1]{\textcolor[rgb]{0.13,0.29,0.53}{#1}}
\newcommand{\DecValTok}[1]{\textcolor[rgb]{0.00,0.00,0.81}{#1}}

\newcommand{\FloatTok}[1]{\textcolor[rgb]{0.00,0.00,0.81}{#1}}

\newcommand{\KeywordTok}[1]{\textcolor[rgb]{0.13,0.29,0.53}{\textbf{#1}}}
\newcommand{\NormalTok}[1]{#1}
\newcommand{\OperatorTok}[1]{\textcolor[rgb]{0.81,0.36,0.00}{\textbf{#1}}}
\newcommand{\OtherTok}[1]{\textcolor[rgb]{0.56,0.35,0.01}{#1}}

\newcommand{\StringTok}[1]{\textcolor[rgb]{0.31,0.60,0.02}{#1}}

\usepackage{longtable,booktabs}
\usepackage{etoolbox}
\makeatletter
\patchcmd\longtable{\par}{\if@noskipsec\mbox{}\fi\par}{}{}
\makeatother
\IfFileExists{footnotehyper.sty}{\usepackage{footnotehyper}}{\usepackage{footnote}}
\makesavenoteenv{longtable}
\usepackage{graphicx}
\makeatletter
\def\maxwidth{\ifdim\Gin@nat@width>\linewidth\linewidth\else\Gin@nat@width\fi}
\def\maxheight{\ifdim\Gin@nat@height>\textheight\textheight\else\Gin@nat@height\fi}
\makeatother
\setkeys{Gin}{width=\maxwidth,height=\maxheight,keepaspectratio}
\makeatletter
\def\fps@figure{htbp}
\makeatother
\providecommand{\tightlist}{%
  \setlength{\itemsep}{0pt}\setlength{\parskip}{0pt}}
\usepackage{hyperref}
\definecolor{linkcolor}{HTML}{D55E00}
\definecolor{citecolor}{HTML}{009E73}
\definecolor{urlcolor}{HTML}{0072B2}
\hypersetup{
    colorlinks,
    linkcolor={linkcolor},
    citecolor={citecolor},
    urlcolor={urlcolor}
}

\makeatletter
\@ifundefined{DecValTok}{}{
  \renewcommand{\DecValTok}[1]{\textcolor[HTML]{009E73}{#1}}
  \renewcommand{\FloatTok}[1]{\textcolor[HTML]{009E73}{#1}}
  
  \renewcommand{\ControlFlowTok}[1]{\textcolor[HTML]{0072B2}{\textbf{#1}}}
  \renewcommand{\OtherTok}[1]{\textcolor[HTML]{000000}{#1}}
  \renewcommand{\CommentTok}[1]{\textcolor[HTML]{999999}{\textit{#1}}}

}
\makeatother

\usepackage{caption}
\usepackage{enumitem}
\setlist[itemize]{topsep=-5pt}
\setlist[enumerate]{topsep=-5pt}

\usepackage[absolute]{textpos}

\usepackage[framemethod=TikZ]{mdframed}
\definecolor{examplecolor}{HTML}{999999}
\surroundwithmdframed[
  leftmargin=\parindent,
  skipabove=\medskipamount,
  skipbelow=\medskipamount,
  linecolor=examplecolor,
  linewidth=1pt,
  roundcorner=5pt
]{example}
\BeforeBeginEnvironment{document}{
  \counterwithout{example}{section}
}

\newcounter{requirement}

\newcommand{\x}{\mathbf{x}}
\newcommand{\w}{\mathbf{w}}
\newcommand{\eps}{\varepsilon}
\newcommand{\Beta}{\operatorname{Beta}}
\newcommand{\Q}{\operatorname{Q}}
\newcommand{\QHD}{\operatorname{Q}_{\operatorname{HD}}}
\newcommand{\QTHD}{\operatorname{Q}_{\operatorname{THD}}}
\newcommand{\HD}{\operatorname{HD}}
\newcommand{\THD}{\operatorname{THD}}
\newcommand{\hf}{{\lfloor h \rfloor}}
\newcommand{\hc}{{\lceil h \rceil}}
\newcommand{\hr}{{\lfloor h \rceil}}
\newcommand{\Exp}{\operatorname{Exp}}
\usepackage{amsmath}
\usepackage{float}
\usepackage{csquotes}
\usepackage{booktabs}
\usepackage{longtable}
\usepackage{array}
\usepackage{multirow}
\usepackage{wrapfig}
\usepackage{colortbl}
\usepackage{pdflscape}
\usepackage{tabu}
\usepackage{threeparttable}
\usepackage{threeparttablex}
\usepackage[normalem]{ulem}
\usepackage{makecell}
\usepackage{xcolor}
\ifxetex
  \usepackage{polyglossia}
  \setmainlanguage[variant=american]{english}
\else
  \usepackage[shorthands=off,main=american]{babel}
\fi
\usepackage[style=alphabetic,sorting=anyt]{biblatex}
\title{Weighted quantile estimators}
\author{Andrey Akinshin\\
JetBrains, \href{mailto:andrey.akinshin@gmail.com}{\nolinkurl{andrey.akinshin@gmail.com}}}
\date{}

\usepackage{amsthm}

\theoremstyle{definition}

\theoremstyle{definition}
\newtheorem{example}{Example}[section]
\theoremstyle{definition}

\theoremstyle{definition}

\theoremstyle{remark}

\begin{document}
\maketitle
\begin{abstract}
In this paper, we consider a generic scheme that allows building
weighted versions of various quantile estimators, such as
traditional quantile estimators based on linear interpolation of two order statistics,
the Harrell--Davis quantile estimator and its trimmed modification.
The obtained weighted quantile estimators are especially useful in the problem
of estimating a distribution at the tail of a time series using quantile exponential smoothing.
The presented approach can also be applied to other problems,
such as quantile estimation of weighted mixture distributions.

\end{abstract}


\hypertarget{sec:intro}{%
\section{Introduction}\label{sec:intro}}

We consider the problem of quantile estimation for a weighted sample.
While this problem arises in different contexts, our primary focus is quantile exponential smoothing.
For the given time series,
we are interested in the quantiles of the distribution at the tail of the series,
which reflects the latest state of the underlying system.
Exponential smoothing suggests assigning weights to sample elements according to the exponential decay law
(the newest measurements get the highest weights, and the oldest measurements get the lowest weights).
A weighted quantile estimator is needed to obtain values of moving quantiles
with a reasonable trade-off between the accuracy of estimations and resistance to obsolete measurements.
Our secondary focus is on the problem of mixture distribution quantile estimation
based on samples from individual distributions.

There are multiple existing weighted quantile estimators, such as
quantile estimator of a weighted mixture distribution based on quantiles of individual samples,
weighted kernel density estimation,
weighted quantile estimations based on a linear combination of two order statistics.
Unfortunately, all of these approaches have limitations,
and they are not always applicable to quantile exponential smoothing.

In this paper, we present a new approach
that allows building weighted versions of existing non-weighted quantile L-estimators,
such as traditional quantile estimators based on a linear combination of two order statistics,
the Harrell--Davis quantile estimator and its trimmed modification.
Our approach is based on a linear combination of multiple order statistics
with linear coefficients obtained from the weights of the sample elements around the target quantile.

The paper is organized as follows.
In Section~\ref{sec:pre},
we review existing approaches to weighted quantile estimations, discuss their disadvantages,
and propose a list of requirements for weighted quantile estimators
that make them practically applicable for quantile exponential smoothing.
In Section~\ref{sec:ess}, we introduce the effective sample size for weighted estimators.
In Sections~\ref{sec:whd} and~\ref{sec:wthd}, we build weighted versions of
the Harrell--Davis quantile estimator and its trimmed modification.
In Section~\ref{sec:whf}, we apply a similar approach to traditional quantile estimators
based on a linear combination of two order statistics and build the corresponding weighted quantile estimators.
In Section~\ref{sec:sim}, we perform a series of simulation studies and
show practical use cases of applying weighted quantile estimators.
In Section~\ref{sec:summary}, we summarize all the results.
In Appendix~\ref{sec:refimpl}, we provide a reference R implementation of the described estimators.

\clearpage

\hypertarget{sec:pre}{%
\section{Preliminaries}\label{sec:pre}}

Let \(\x = \{ x_1, x_2, \ldots, x_n \}\) be a sample of size \(n\).
We assign non-negative weight coefficients \(w_i\) with a positive sum for all sample elements:

\[
\w = \{ w_1, w_2, \ldots, w_n \}, \quad w_i \geq 0, \quad \sum_{i=1}^{n} w_i > 0.
\]

For simplification, we also consider normalized (standardized) weights \(\overline{\w}\):

\[
\overline{\w} = \{ \overline{w}_1, \overline{w}_2, \ldots, \overline{w}_n \}, \quad
  \overline{w}_i = \frac{w_i}{\sum_{i=1}^{n} w_i}.
\]

Let \(x_{(i)}\) be the \(i^\textrm{th}\) order statistic of \(\x\),
\(w_{(i)}\) be the weight associated with \(x_{(i)}\),
and \(\overline{w}_{(i)}\) be the corresponding normalized weight.
We denote partial sums of \(w_{(i)}\) and \(\overline{w}_{(i)}\) by \(s_i(\cdot)\):

\[
s_i(\w) = \sum_{j=1}^{i} w_{(j)}, \quad
s_i(\overline{\w}) = \sum_{j=1}^{i} \overline{w}_{(j)}, \quad
s_0(\w) = s_0(\overline{\w}) = 0.
\]

In the non-weighted case, we can consider a quantile estimator \(\Q(\x, p)\)
that estimates the \(p^\textrm{th}\) quantile of the underlying distribution.
We want to build a weighted quantile estimator \(\Q^*(\x, \w, p)\)
so that we can estimate the quantiles of a weighted sample.
The problem of weighted quantile estimations arises in different contexts.

One of the possible applications is estimating the quantiles of a mixture distribution.
Let us consider an example of building such a weighted estimator
for a mixture of three distributions given by their cumulative distribution functions (CDFs)
\(F_X\), \(F_Y\), and \(F_Z\) with weights \(w_X\), \(w_Y\), and \(w_Z\).
The weighted mixture is given by \(F=\overline{w}_X F_X + \overline{w}_Y F_Y + \overline{w}_Z F_Z\).
Let us say that we have samples \(\mathbf{x}\), \(\mathbf{y}\), and \(\mathbf{z}\) from \(F_X\), \(F_Y\), and \(F_Z\);
and we want to estimate the quantile function \(F^{-1}\) of the mixture distribution \(F\).
If each sample contains a sufficient number of elements, we can consider a straightforward approach:

\begin{enumerate}
\def\labelenumi{\arabic{enumi}.}
\tightlist
\item
  Obtain empirical distribution quantile functions \(\hat{F}^{-1}_X\), \(\hat{F}^{-1}_Y\), \(\hat{F}^{-1}_Z\)
  based on the given samples;
\item
  Invert quantile functions and obtain estimations \(\hat{F}_X\), \(\hat{F}_Y\), \(\hat{F}_Z\)
  of the CDFs for each distribution;
\item
  Combine these CDFs and build an estimation
  \(\hat{F}=\overline{w}_X\hat{F}_X+\overline{w}_Y\hat{F}_Y+\overline{w}_Z\hat{F}_Z\) of the mixture CDF;
\item
  Invert \(\hat{F}\) and get the estimation \(\hat{F}^{-1}\) of the mixture distribution quantile function.
\end{enumerate}

Unfortunately, this scheme performs poorly in the case of small sample sizes
due to inaccurate estimations of the individual CDFs.
Also, it is not extendable to the exponential smoothing problem.

Another approach that can be considered is the weighted kernel density estimation (KDE).
It suggests estimating the probability density function (PDF) \(f\) as follows:

\[
\hat{f}(x) = \sum_{i=1}^n \frac{\overline{w}_i}{h} K \left( \frac{x - x_i}{h} \right),
\]

where \(K\) is the kernel (typically, the standard normal distribution is used),
\(h\) is the bandwidth.
Next, we obtain the corresponding CDF by integrating the PDF and get the quantile function by inverting the CDF.
This approach is applicable in some cases, but it inherits all the disadvantages of the non-weighted KDE.
Firstly, it heavily depends on the choice of the kernel and the bandwidth.
For example, Silverman's and Scott's rules of thumb for bandwidth selection
(which are used by default in many statistical packages)
perform poorly in the non-normal case and mask multimodality.
Secondly, it extends the range of the quantile values:
lower and higher quantiles exceed the minimum and maximum values of \(\x\), which is not always acceptable.
Still, there are some implementations of weighted KDE that use various bandwidth and kernel selectors
(e.g., see \autocite{wolters2018,wang2007,guillamn1984}).

The proper choice of a weighted quantile estimator depends not only on the estimator properties,
but also on the research goals.
Different goals require different estimators.
We already discussed the problem of estimating the quantiles of a mixture distribution based on individual samples.
This paper primarily focuses on another problem called \emph{quantile exponential smoothing}.
Within this problem, we consider \(\x\) as a time series of measurements.
The goal is to estimate the distribution at the tail of this time series
(the actual or latest state of the underlying system).
The latest series element \(x_n\) is the most actual one, but we cannot build a distribution based on a single element.
Therefore, we have to take into account more elements at the end of \(\x\).
However, if we take too many elements, we may corrupt the estimations due to obsolete measurements.
This problem is illustrated in Example~\ref{exm:smoothing-problem}.

\begin{example}
\protect\hypertarget{exm:smoothing-problem}{}\label{exm:smoothing-problem}Let us consider the problem of estimating distribution quantiles at the tail of a time series.
One of the simplest approaches is to take the last \(k\) measurements without weights
and estimate quantiles based on the obtained data.
This approach may lead to inaccurate estimations.
Let us illustrate it using two timeline plots presented in Figure~\ref{fig:smoothing-problem}.

\begin{figure}[H]

{\centering \includegraphics{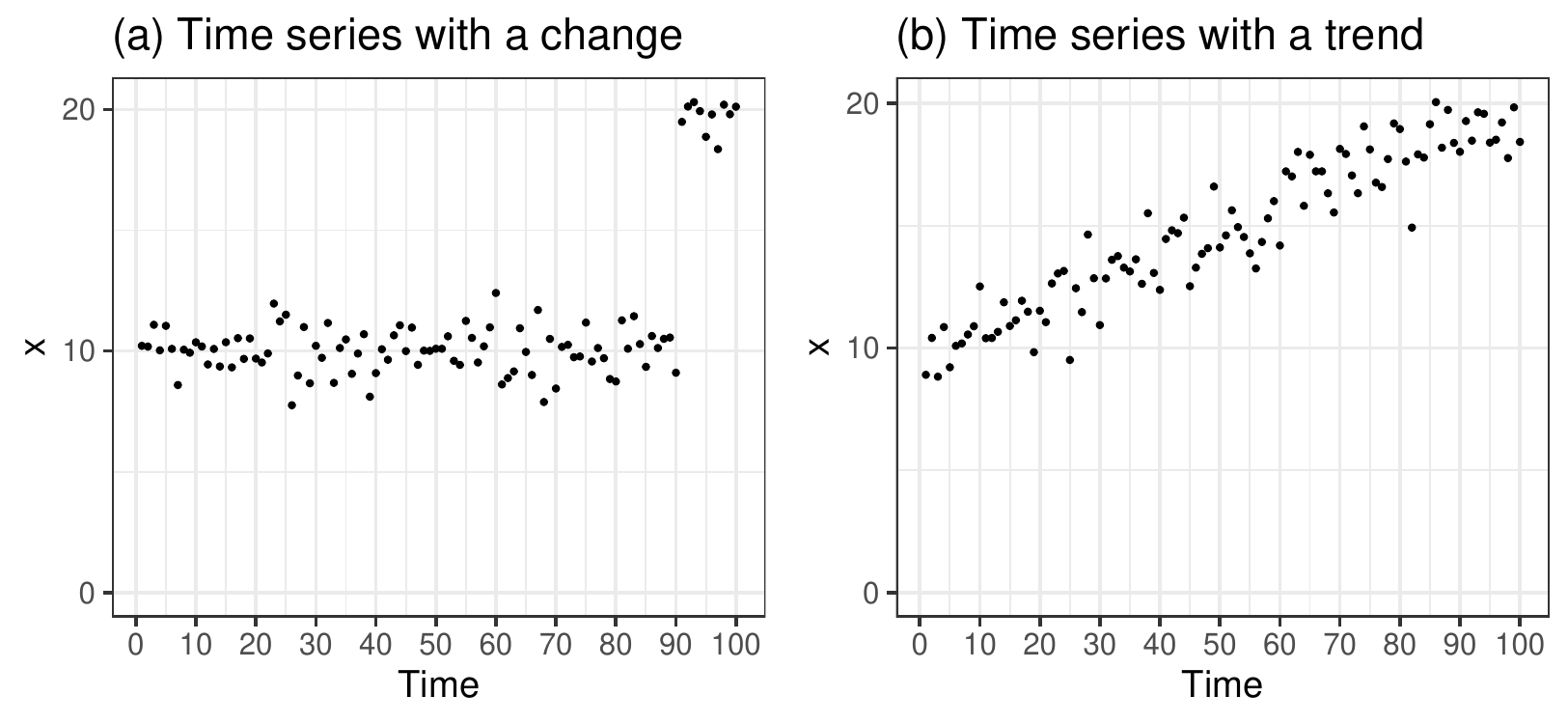} 

}

\caption{Two timeline plots.}\label{fig:smoothing-problem}
\end{figure}

In Figure~\ref{fig:smoothing-problem}a, we can see a change point after the first 90 measurements.
If the number of considered measurements \(k\) is greater than 10,
the selected elements include ``obsolete'' values which corrupt the quantile estimations
(especially lower quantiles).
If \(k\) is less than 10, we lose statistical efficiency due to an insufficiently large sample size.
A possible solution for this problem is to automatically detect change points in the time series
(an overview of change point detectors can be found in \autocite{truong2020}),
and omit all the data before the last change point.

A drawback of this approach is presented in Figure~\ref{fig:smoothing-problem}b.
Here we can see a trend: the time series values are constantly increasing without any obvious change points.
How should we choose the optimal value of \(k\) in this situation?
Small values of \(k\) prevent us from having enough data, which is essential for accurate estimations.
Large values of \(k\) introduce too many obsolete values,
contaminating the selected subsample and corrupting the estimations.
Thus, we have a trade-off between the accuracy of estimations and resistance to obsolete measurements.

This trade-off is a severe issue in any approach that estimates the distribution at the tail of the time series
using non-weighted subsamples formed from the last \(k\) measurements.
Adaptive strategies of choosing \(k\) based on change point detectors
slightly reduce the risk of obtaining invalid estimations,
but they have smoothing issues when \(x_{n-k}\) is around a change point.
\end{example}

\clearpage

The described problem can be mitigated using exponential smoothing.
This approach assumes that we assign exponentially decreasing weights to the sample elements.
That looks reasonable: the older the measurement, the lower its impact on the final estimations.
The idea of exponential smoothing is widely used for the arithmetic mean
(the corresponding approach is known as the exponentially weighted moving average).
However, the mean is not robust: a single extreme outlier can corrupt the mean estimations.
That is why it makes sense to switch to the quantiles,
which allow us to describe the whole distribution even in the non-parametric heavy-tailed case.
In order to apply exponential smoothing, we need weighted quantile estimators.

It is important to understand the difference between
the problem of estimating mixture distribution quantiles and the smoothing problem.
It is worth noting that this difference usually appears when the middle part of the estimated distribution
contains low-density regions, which often arise due to multimodality.
Such regions are always a source of trouble in quantile estimation
since we typically do not have enough data for accurate estimations.
This issue is illustrated in Example~\ref{exm:smoothing-vs-mixture}.

\begin{example}
\protect\hypertarget{exm:smoothing-vs-mixture}{}\label{exm:smoothing-vs-mixture}Let us consider the following distribution:

\[
F =
  \eps F_{\delta_{-1000}} +
  \frac{1-3\eps}{2} F_{\mathcal{U}(0,1)} +
  \eps F_{\delta_{10}} +
  \frac{1-3\eps}{2} F_{\mathcal{U}(99,100)} +
  \eps F_{\delta_{1000}},
\]

where \(F_{\star}\) is a CDF of the given distribution,
\(\delta_t\) is the Dirac delta distribution (it has all its mass at \(t\)),
\(\mathcal{U}(a, b)\) is a continuous uniform distribution on \([a;b]\),
\(\eps\) is a small positive constant.
When we consider the problem of obtaining the true quantile values of the mixture distribution \(F\),
we should expect

\[
F^{-1}(0) = -1000,\quad F^{-1}(0.5) = 10,\quad F^{-1}(1) = 1000,
\]

which matches the true quantile values of \(F\) for any positive \(\eps\).

However, in the smoothing problem, we want to have a negligible impact of
\(F_{\delta_{-1000}}\), \(F_{\delta_{10}}\), \(F_{\delta_{1000}}\) on the quantile estimations when \(\eps \to 0\).
More specifically, we want to get

\[
F^{-1}(0) \approx 0,\quad F^{-1}(0.5) \approx 50,\quad F^{-1}(1) \approx 100.
\]

Although the expected values of \(F^{-1}(p)\) are typically quite similar in both problems for most values of \(p\),
it is important to keep in mind the corner cases.
\end{example}

In this paper, we primarily focus on the problem of quantile exponential smoothing.
To make the weighted quantile estimators practically useful, we define a list of desired properties
that are expressed in the form of tree Requirements (\ref{req:consistency},~\ref{req:zero},~\ref{req:stability}).

\textbf{Requirement R1: consistency with existing quantile estimators.} \refstepcounter{requirement}\label{req:consistency}
There are multiple ways to estimate non-weighted quantiles.
For example, we can consider
the traditional quantile estimators based on linear interpolation of two order statistics (see \autocite{hyndman1996}),
the Harrell--Davis quantile estimator (see \autocite{harrell1982}),
and its trimmed modification (see \autocite{akinshin2022thdqe}).
Different estimators have different characteristics in terms of
statistical efficiency, computational efficiency, and robustness.
Instead of creating a family of new weighted quantile estimators with different sets of properties,
we want to build a generalization of the existing non-weighted estimators and inherit their properties.
The generalized weighted estimator should be consistent with the original non-weighted estimator
on the unit vector of weights \(\w = \{ 1, 1, \ldots, 1 \}\):

\[
\Q^*(\x, \{ 1, 1, \ldots, 1 \}, p) = \Q(\x, p).
\]

\textbf{Requirement R2: zero weight support.} \refstepcounter{requirement}\label{req:zero}
It is also reasonable to require that sample elements with zero weights should not affect the estimation:

\[
\Q^*(\{x_1, x_2, \ldots, x_{n-1}, x_n \}, \{w_1, w_2, \ldots, w_{n-1}, 0\}, p) =
\Q^*(\{x_1, x_2, \ldots, x_{n-1} \}, \{w_1, w_2, \ldots, w_{n-1}\}, p).
\]

\textbf{Requirement R3: stability.} \refstepcounter{requirement}\label{req:stability}
When we use exponential smoothing, we have to define the smoothing factor.
This factor often requires some adjustments in order to achieve a balance between statistical efficiency and robustness.
It is reasonable to expect
that small changes in the smoothing factor should not produce significant changes in the obtained estimations.
The continuity of the quantile estimations with respect to the weight coefficient makes
the adjustment process more simple and controllable.
Another essential use case is the addition of new elements,
which leads to a slight reduction of weights of the existing elements.
It is desirable that such a small reduction will not produce significant changes in the estimation.
Generalizing, we require that minor changes in weight coefficients should not produce a major impact on the estimation.
More formally,

\[
\lim_{\eps_i \to 0} \Q^*(\{x_1, x_2, \ldots, x_n \}, \{w_1 + \eps_1, w_2 + \eps_2, \ldots, w_n + \eps_n \}, p) \to
  \Q^*(\{x_1, x_2, \ldots, x_n \}, \{w_1, w_2, \ldots, w_n\}, p).
\]

This rule has a practically interesting use case of dropping elements with small weights.
Indeed, it is reasonable to build a weighted sample for exponential smoothing based not on all available data
but only on the last \(k\) elements.
With the exponential decay law, old elements get extremely small weights and
therefore should not produce a noticeable impact on the estimation.
In order to speed up the calculations, we should have an opportunity to exclude such elements from the sample:

\[
\begin{split}
\lim_{\eps \to 0} & \Q^*(\{ x_1, x_2, \ldots, x_{n-1}, x_n \}, \{ \eps, w_2, \ldots, w_{n-1}, w_n \}) \stackrel{R3}{\to}\\
  \stackrel{R3}{\to} & \Q^*(\{ x_1, x_2, \ldots, x_{n-1}, x_n \}, \{ 0, w_2, \ldots, w_{n-1}, w_n \}) \stackrel{R2}{=}\\
  \stackrel{R2}{=} & \Q(\{ x_2, x_3, \ldots, x_n \}, \{ w_2, w_3, \ldots, w_n \})
\end{split}
\]

\bigskip

Now let us explore some of the existing solutions for the weighted quantile estimators in popular statistical packages.
Most of these solutions
are based on a single order statistic or a linear interpolation of two order statistics.
Approaches based on a single order statistic violate Requirement~\ref{req:stability}:
small fluctuations in \(\w\) around the threshold point
can ``switch'' the estimation from \(x_{(i)}\) to \(x_{(i-1)}\) or \(x_{(i+1)}\).
Approaches based on a linear interpolation of two subsequent order statistics
also violate Requirement~\ref{req:stability} as shown in Example~\ref{exm:linear-issue1}.

\begin{example}
\protect\hypertarget{exm:linear-issue1}{}\label{exm:linear-issue1}\(\x = \{ 0, 1, 1, 100 \}\), \(\w = \{ 1, \eps, \eps, 1 \}\), \(p = 0.5\).\\
When \(\eps=0\), \(x_2\) and \(x_3\) have zero weights and should be omitted,
which gives us two equally weighted elements \(\{ 0, 100 \}\).
The traditional sample median of this sample is \(50\).
When \(\eps>0\), the median estimation based on two subsequent order statistics
should use \(x_2\) and \(x_3\) due to the symmetry of \(\w\).
Since \(x_2=x_3=1\), the median estimation will be equal to \(1\).
Thus, the transition from \(\eps=0\) to \(\eps \to +0\) switches the median estimation from \(50\) to \(1\),
which is a violation of Requirement~\ref{req:stability}.
\end{example}

Sometimes, approaches based on a linear interpolation of two non-subsequent order statistics are used in order
to work around the problem from Example~\ref{exm:linear-issue1}.
However, it is always possible to find examples that show violations of Requirement~\ref{req:stability}.
Instead of analyzing all possible linear interpolation equations and discussing their disadvantages,
we briefly provide examples of similar violations in popular weighted quantile estimator implementations.
We limit our consideration scope to the R language
since it is one of the most popular programming languages for statistical computing.
Let us review the popular R implementation of weighted quantiles
from the following CRAN\footnote{\url{https://cran.r-project.org/}} packages:
\texttt{modi\ 0.1.0} (\autocite{modi}),
\texttt{laeken\ 0.5.2} (\autocite{laeken}),
\texttt{MetricsWeighted\ 0.5.4} (\autocite{MetricsWeighted}),
\texttt{spatstat.geom\ 2.4-0} (\autocite{spatstat}),
\texttt{matrixStats\ 0.62.0} (\autocite{matrixStats}),
\texttt{DescTools\ 0.99.46} (\autocite{desctools}),
\texttt{Hmisc\ 4.7-1} (\autocite{Hmisc}).
Undesired behavior patterns are presented in Example~\ref{exm:linear-issue2} and Example~\ref{exm:linear-issue3}.

\clearpage

\begin{example}
\protect\hypertarget{exm:linear-issue2}{}\label{exm:linear-issue2}\(\x = \{ 0, 1, 100 \}\), \(\w_A = \{ 1, 0, 1 \}\), \(\w_B = \{ 1, 0.00001, 1 \}\), \(p = 0.5\).\\
The only difference between \(\w_A\) and \(\w_B\) is in the second element: it changes from \(0\) to \(0.00001\).
Since the change is small, we can expect a small difference between \(\Q(\x, \w_A, p)\) and \(\Q(\x, \w_B, p)\).
Let us check the actual estimation values using \texttt{modi}, \texttt{laeken}, \texttt{MetricsWeighted}, \texttt{spatstat.geom}, \texttt{matrixStats}.

\begin{Shaded}
\begin{Highlighting}[]
\NormalTok{x \textless{}{-}}\StringTok{ }\KeywordTok{c}\NormalTok{(}\DecValTok{0}\NormalTok{, }\DecValTok{1}\NormalTok{, }\DecValTok{100}\NormalTok{)}
\NormalTok{wA \textless{}{-}}\StringTok{ }\KeywordTok{c}\NormalTok{(}\DecValTok{1}\NormalTok{, }\FloatTok{0.00000}\NormalTok{, }\DecValTok{1}\NormalTok{)}
\NormalTok{wB \textless{}{-}}\StringTok{ }\KeywordTok{c}\NormalTok{(}\DecValTok{1}\NormalTok{, }\FloatTok{0.00001}\NormalTok{, }\DecValTok{1}\NormalTok{)}
\KeywordTok{message}\NormalTok{(modi}\OperatorTok{::}\KeywordTok{weighted.quantile}\NormalTok{(x, wA, }\FloatTok{0.5}\NormalTok{), }\StringTok{" | "}\NormalTok{,}
\NormalTok{        modi}\OperatorTok{::}\KeywordTok{weighted.quantile}\NormalTok{(x, wB, }\FloatTok{0.5}\NormalTok{))}
\CommentTok{\#\# 100 | 1}
\KeywordTok{message}\NormalTok{(laeken}\OperatorTok{::}\KeywordTok{weightedQuantile}\NormalTok{(x, wA, }\FloatTok{0.5}\NormalTok{), }\StringTok{" | "}\NormalTok{,}
\NormalTok{        laeken}\OperatorTok{::}\KeywordTok{weightedQuantile}\NormalTok{(x, wB, }\FloatTok{0.5}\NormalTok{))}
\CommentTok{\#\# 100 | 1}
\KeywordTok{message}\NormalTok{(MetricsWeighted}\OperatorTok{::}\KeywordTok{weighted\_quantile}\NormalTok{(x, wA, }\FloatTok{0.5}\NormalTok{), }\StringTok{" | "}\NormalTok{,}
\NormalTok{        MetricsWeighted}\OperatorTok{::}\KeywordTok{weighted\_quantile}\NormalTok{(x, wB, }\FloatTok{0.5}\NormalTok{))}
\CommentTok{\#\# 100 | 1}
\KeywordTok{message}\NormalTok{(spatstat.geom}\OperatorTok{::}\KeywordTok{weighted.quantile}\NormalTok{(x, wA, }\FloatTok{0.5}\NormalTok{), }\StringTok{" | "}\NormalTok{,}
\NormalTok{        spatstat.geom}\OperatorTok{::}\KeywordTok{weighted.quantile}\NormalTok{(x, wB, }\FloatTok{0.5}\NormalTok{))}
\CommentTok{\#\# 1 | 0.499999999994449}
\KeywordTok{message}\NormalTok{(matrixStats}\OperatorTok{::}\KeywordTok{weightedMedian}\NormalTok{(x, wA), }\StringTok{" | "}\NormalTok{,}
\NormalTok{        matrixStats}\OperatorTok{::}\KeywordTok{weightedMedian}\NormalTok{(x, wB))}
\CommentTok{\#\# 50 | 1.00000000000003}
\end{Highlighting}
\end{Shaded}

As we can see, all the considered packages have a discontinuity in quantile estimations around \(w_2 = 0\).
It is trivial to formally prove this fact for each particular implementation.
\end{example}

\begin{example}
\protect\hypertarget{exm:linear-issue3}{}\label{exm:linear-issue3}\(\x = \{ 0, 1, 100 \}\), \(\w_C = \{ 1, 0.99999, 1 \}\), \(\w_D = \{ 1, 1, 1 \}\), \(p = 0.5\).\\
This case is similar to Example~\ref{exm:linear-issue2}, but now we change \(w_2\) from \(0.99999\) to \(1\).
Let us review the actual estimation values using \texttt{DescTools} and \texttt{Hmisc}.

\begin{Shaded}
\begin{Highlighting}[]
\NormalTok{x \textless{}{-}}\StringTok{ }\KeywordTok{c}\NormalTok{(}\DecValTok{0}\NormalTok{, }\DecValTok{1}\NormalTok{, }\DecValTok{100}\NormalTok{)}
\NormalTok{wC \textless{}{-}}\StringTok{ }\KeywordTok{c}\NormalTok{(}\DecValTok{1}\NormalTok{, }\FloatTok{0.99999}\NormalTok{, }\DecValTok{1}\NormalTok{)}
\NormalTok{wD \textless{}{-}}\StringTok{ }\KeywordTok{c}\NormalTok{(}\DecValTok{1}\NormalTok{, }\FloatTok{1.00000}\NormalTok{, }\DecValTok{1}\NormalTok{)}
\KeywordTok{message}\NormalTok{(DescTools}\OperatorTok{::}\KeywordTok{Quantile}\NormalTok{(x, wC, }\FloatTok{0.5}\NormalTok{), }\StringTok{" | "}\NormalTok{,}
\NormalTok{        DescTools}\OperatorTok{::}\KeywordTok{Quantile}\NormalTok{(x, wD, }\FloatTok{0.5}\NormalTok{))}
\CommentTok{\#\# 99.9995 | 1}
\KeywordTok{message}\NormalTok{(Hmisc}\OperatorTok{::}\KeywordTok{wtd.quantile}\NormalTok{(x, wC, }\FloatTok{0.5}\NormalTok{), }\StringTok{" | "}\NormalTok{,}
\NormalTok{        Hmisc}\OperatorTok{::}\KeywordTok{wtd.quantile}\NormalTok{(x, wD, }\FloatTok{0.5}\NormalTok{))}
\CommentTok{\#\# 99.9995 | 1}
\end{Highlighting}
\end{Shaded}

As we can see, these packages have a discontinuity around \(w_2 = 1\).
Moreover, the \(\Q(\x, \w_C, p)\) estimations are counterintuitive.
Indeed, since we expect \(\Q^*(\{ 0, 1, 100 \}, \{ 1, 0, 1 \}, 0.5) = 50\) (Requirement~\ref{req:zero})
and \(\Q^*(\{ 0, 1, 100 \}, \{ 1, 1, 1 \}, 0.5) = 1\) (Requirement~\ref{req:consistency}),
it is also reasonable to expect that \(\Q^*(\{ 0, 1, 100 \}, \{ 1, 0.99999, 1 \}, 0.5) \in [1; 50]\).
However, we observe a strange estimation value of \(99.9995\) for both packages.
\end{example}

Thus, we cannot rely on existing weighted quantile estimator implementations and
we cannot reuse approaches for weighted mixture distribution.
Therefore, we will build a new scheme for building weighted versions of some existing quantile estimators.

\clearpage

\hypertarget{sec:ess}{%
\section{Effective sample size}\label{sec:ess}}

All kinds of non-weighted quantile estimators explicitly or implicitly use the sample size.
In order to satisfy Requirement~\ref{req:zero}, the sample size needs adjustments.
Indeed, if we add another sample element with zero weight, we increment the actual sample size by one,
which should not lead to changes in the estimations.
To solve this problem, we suggest using Kish's effective sample size (see \autocite{kish1965}) given by:

\begin{equation}
n^*(\w) =
  \frac{\left( \sum_{i=1}^n w_i \right)^2}{\sum_{i=1}^n w_i^2 } =
  \frac{1}{\sum_{i=1}^n \overline{w}_i^2 }.
\label{eq:kish}
\end{equation}

To better understand Kish's effective sample size, see Example~\ref{exm:kish}.

\begin{example}
\protect\hypertarget{exm:kish}{}\label{exm:kish}

Here are some examples of Kish's effective sample size:

\begin{enumerate}
\def\labelenumi{\arabic{enumi}.}
\tightlist
\item
  \(n^*(\{1, 1, 1\}) = 3\),
\item
  \(n^*(\{2, 2, 2\}) = 3\),
\item
  \(n^*(\{1, 1, 1, 0, 0\}) = 3\),
\item
  \(n^*(\{1, 1, 1, 0.00001\}) \approx 3.00002\),
\item
  \(n^*(\{1, 2, 3, 4, 5\}) \approx 4.090909\).
\end{enumerate}

\end{example}

We can also consider the Huggins--Roy family of effective sample sizes
(proposed in \autocite{huggins2019}, advocated in \autocite{elvira2022}):

\[
\operatorname{ESS}_\beta(\overline{\mathbf{w}}) = \begin{cases}
  n - {\#}(\overline{w}_i = 0), & \textrm{if}\; \beta = 0, \\
  \exp \left( -\sum_{i=1}^n \overline{w}_i \log \overline{w}_i \right), & \textrm{if}\; \beta = 1, \\
  \frac{1}{\max[\overline{w}_1, \overline{w}_2, \ldots, \overline{w}_n]}, & \textrm{if}\; \beta = \infty, \\
  \left( \frac{1}{\sum_{i=1}^n \overline{w}_i^\beta } \right)^{\frac{1}{\beta - 1}} =
  \left( \sum_{i=1}^n \overline{w}_i^\beta \right)^{\frac{1}{1 - \beta}}, & \textrm{otherwise}.
\end{cases}
\]

All estimators in this family satisfy Requirement~\ref{req:zero}:
additional zero weights do not affect the effective sample size value.
It is also easy to see that Kish's Equation~\eqref{eq:kish} is a specific case of the Huggins--Roy family:
\(n^* = \operatorname{ESS}_2\).
We continue building weighted quantile estimators using Kish's approach,
but other effective sample size estimators can be considered as well.

\hypertarget{sec:whd}{%
\section{Weighted Harrell--Davis quantile estimator}\label{sec:whd}}

The Harrell--Davis quantile estimator (see \autocite{harrell1982}) evaluates quantiles
as a linear combination of all order statistics.
Although this approach is not robust (its breakdown point is zero),
it provides higher statistical efficiency in the cases of light-tailed distributions,
which makes it a practically reasonable option under lighttailedness.
We start with this particular estimator instead of the traditional approaches because
it has an intuition-friendly weighted case generalization.

The classic non-weighted Harrell--Davis quantile estimator is given by:

\begin{equation}
\QHD(\x, p) = \sum_{i=1}^{n} W_{\operatorname{HD},i} \cdot x_{(i)},\quad
W_{\operatorname{HD},i} = I_{t_i}(\alpha, \beta) - I_{t_{i-1}}(\alpha, \beta),
\label{eq:hd}
\end{equation}

where \(I_t(\alpha, \beta)\) is the regularized incomplete beta function,
\(t_i = i/n\), \(\alpha = (n+1)p\), \(\;\beta = (n+1)(1-p)\).
See Example~\ref{exm:hd-intro} to get the intuition behind the Harrell--Davis quantile estimator.

\clearpage

\begin{example}
\protect\hypertarget{exm:hd-intro}{}\label{exm:hd-intro}\(\x = \{ 1, 2, 4, 8, 16 \}\), \(p = 0.5\).\\
We estimate the median for a sample of size \(5\),
which gives us \(\alpha = \beta = 3\).
Let us consider the PDF of the beta distribution
\(\Beta(3, 3)\) presented in Figure~\ref{fig:hd-intro}.

\begin{figure}[H]

{\centering \includegraphics{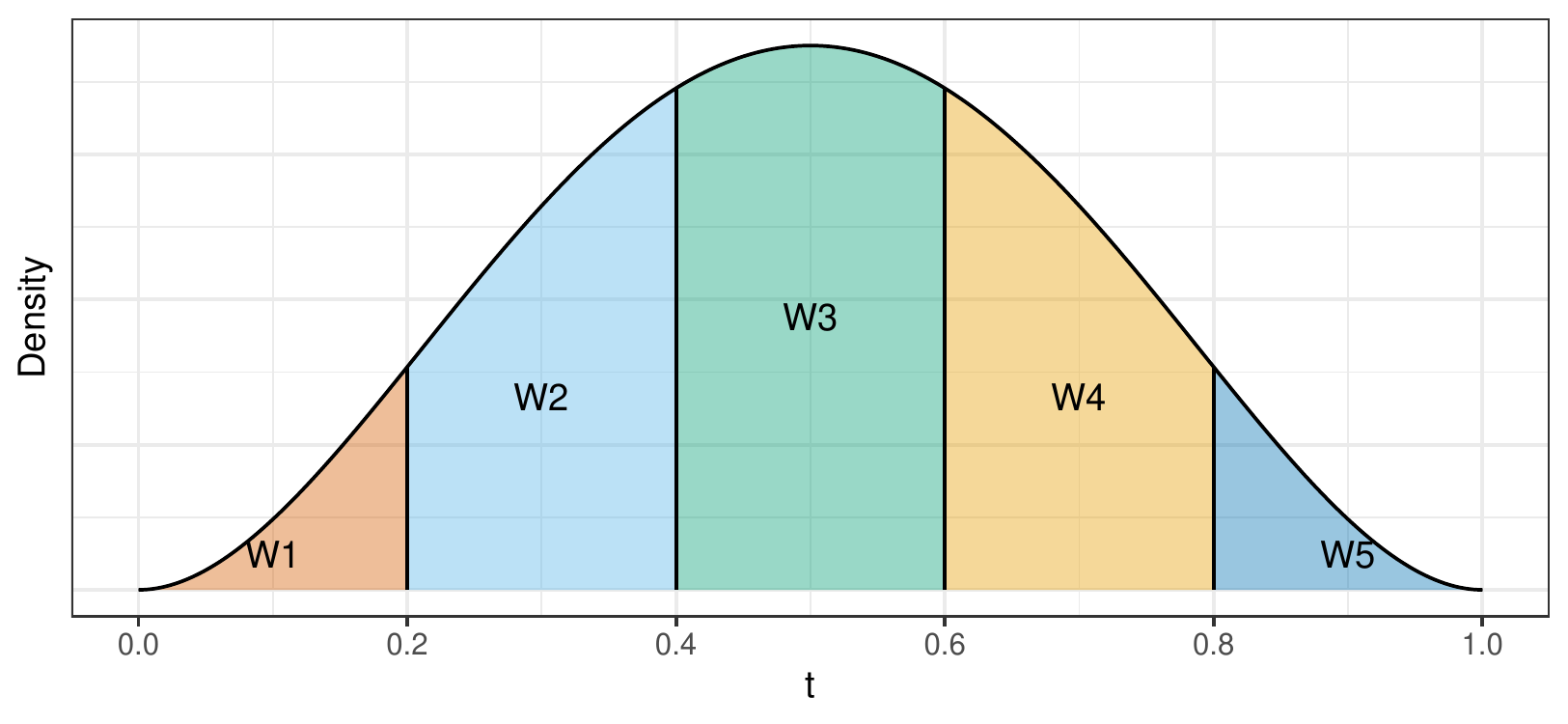} 

}

\caption{Beta(3, 3) probability density function.}\label{fig:hd-intro}
\end{figure}

This PDF is split into five fragments with cut points given by

\[
t_{\{ 0..5\}} = \{ t_0, t_1, t_2, t_3, t_4, t_5 \} = \{ 0, 0.2, 0.4, 0.6, 0.8, 1 \}.
\]

Since \(I_t(\alpha, \beta)\) is the CDF of the beta distribution \(\Beta(\alpha, \beta)\),
the area of the \(i^\textrm{th}\) fragment is exactly
\(I_{t_i}(\alpha, \beta) - I_{t_{i-1}}(\alpha, \beta) = W_{\operatorname{HD},i}\).
Here are the corresponding linear coefficient values:

\[
W_{\operatorname{HD},\{1..5\}} \approx \{ 0.058, 0.26, 0.365, 0.26, 0.058 \}.
\]

Now we can calculate the Harrell--Davis median estimation:

\[
\QHD(\x, p) = \sum_{i=1}^{n} W_{\operatorname{HD},i} \cdot x_{(i)} \approx 5.04.
\]

We can interpret the \(W_{\operatorname{HD},i}\) coefficients
as probabilities of observing the target quantile at the given position.
Thus, the Harrell--Davis estimation is a weighted sum of all order statistics according to these probabilities.
\end{example}

In the non-weighted case, all segments \([t_{i-1}; t_i]\) have the same width of \(1/n\).
For the weighted case, we suggest replacing them with intervals \([t^*_{i-1}; t^*_i]\)
that have widths proportional to the element weights \(w_{(i)}\).
Thus, we can define \(t^*_i\) via partial sums of normalized weight coefficients:

\begin{equation}
t^*_i = s_i(\overline{\w}).
\label{eq:whd-t}
\end{equation}

Since we use Kish's effective sample size \(n^*\) in the weighted case,
the values of \(\alpha\) and \(\beta\) should be also properly adjusted:

\begin{equation}
\alpha^* = (n^* + 1) p, \quad \beta^* = (n^* + 1) (1 - p).
\label{eq:whd-ab}
\end{equation}

Using~\eqref{eq:whd-t} and~\eqref{eq:whd-ab},
we can define the weighted Harrell--Davis quantile estimator:

\begin{equation}
\QHD^*(\x, \w, p) = \sum_{i=1}^{n} W^*_{\HD,i} \cdot x_{(i)},\quad
W^*_{\HD,i} = I_{t^*_i}(\alpha^*, \beta^*) - I_{t^*_{i-1}}(\alpha^*, \beta^*).
\label{eq:whd}
\end{equation}

The only corner cases for the Harrell--Davis quantile estimator are \(p=0\) and \(p=1\)
since \(\Beta(\alpha, \beta)\) is defined only when \(\alpha, \beta > 0\).
For all \(p \in (0;1)\), \(\QHD^*(\x, \w, p)\) satisfies all of the declared requirements.

The usage of \(\QHD^*(\x, \w, p)\) is demonstrated in Example~\ref{exm:whd1} and Example~\ref{exm:whd2}.

\begin{example}
\protect\hypertarget{exm:whd1}{}\label{exm:whd1}\(\x = \{ 1, 2, 3, 4, 5 \}\), \(\w = \{ 1, 1, 0, 0, 1 \}\), \(p = 0.5\).

Without the given weights \(\w\),
the linear coefficients \(W_{\HD,i}\) are defined as shown in Example~\ref{exm:hd-intro}.
When we consider the weighted case, \(x_{(3)}\) and \(x_{(4)}\) should be automatically omitted because \(w_{(3)}=w_{(4)}=0\).
From~\eqref{eq:kish} and~\eqref{eq:whd-ab}, we get \(n^* = 3,\, \alpha^* = \beta^* = 2\),
which gives us \(\Beta(2, 2)\) (the PDF is shown in Figure~\ref{fig:whd1}).
The cut points are given by:

\[
t^*_{\{0..5\}} = \{ 0,\; 1/3,\; 2/3,\; 2/3,\; 2/3,\; 1 \}.
\]

\begin{figure}[H]

{\centering \includegraphics{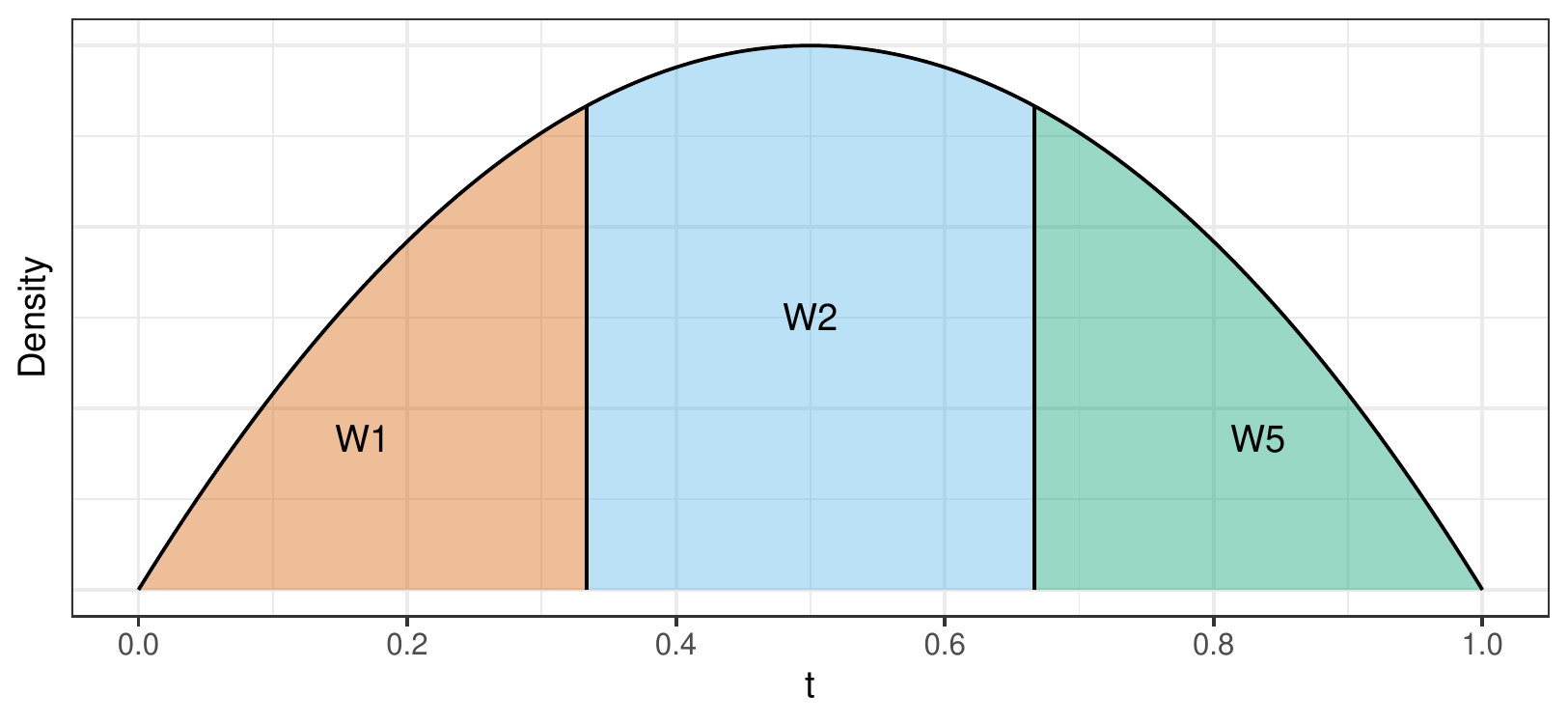} 

}

\caption{Beta(2, 2) probability density function.}\label{fig:whd1}
\end{figure}

Using \(I_t(2, 2)\), we can obtain the Harrell--Davis linear coefficients:

\[
W^*_{\operatorname{HD},\{1..5\}} \approx \{ 0.259, 0.481, 0, 0, 0.259 \}.
\]

Thus, elements \(x_{(3)}\) and \(x_{(4)}\) are indeed automatically eliminated.
The median is estimated as in the non-weighted case for three elements:

\[
\QHD^*(\{ 1, 2, 3, 4, 5 \}, \{ 1, 0, 0, 1, 1 \}, 0.5) = \QHD(\{ 1, 2, 5 \}, 0.5) \approx 2.518519.
\]
\end{example}

\clearpage

\begin{example}
\protect\hypertarget{exm:whd2}{}\label{exm:whd2}\(\x = \{ 1, 2, 3, 4, 5 \}\), \(\w = \{ 0.4, 0.4, 0.05, 0.05, 0.1 \}\), \(p = 0.5\).\\
From~\eqref{eq:kish} and~\eqref{eq:whd-ab}, we get

\[
n^* \approx 2.985,\quad
\alpha^* = \beta^* \approx 1.993,
\]

which gives us \(\Beta(1.993, 1.993)\)
(the PDF is shown in Figure~\ref{fig:whd2}).
The cut points are given by:

\[
t^*_{\{0..5\}} = \{ 0, 0.4, 0.8, 0.85, 0.9, 1 \}.
\]

\begin{figure}[H]

{\centering \includegraphics{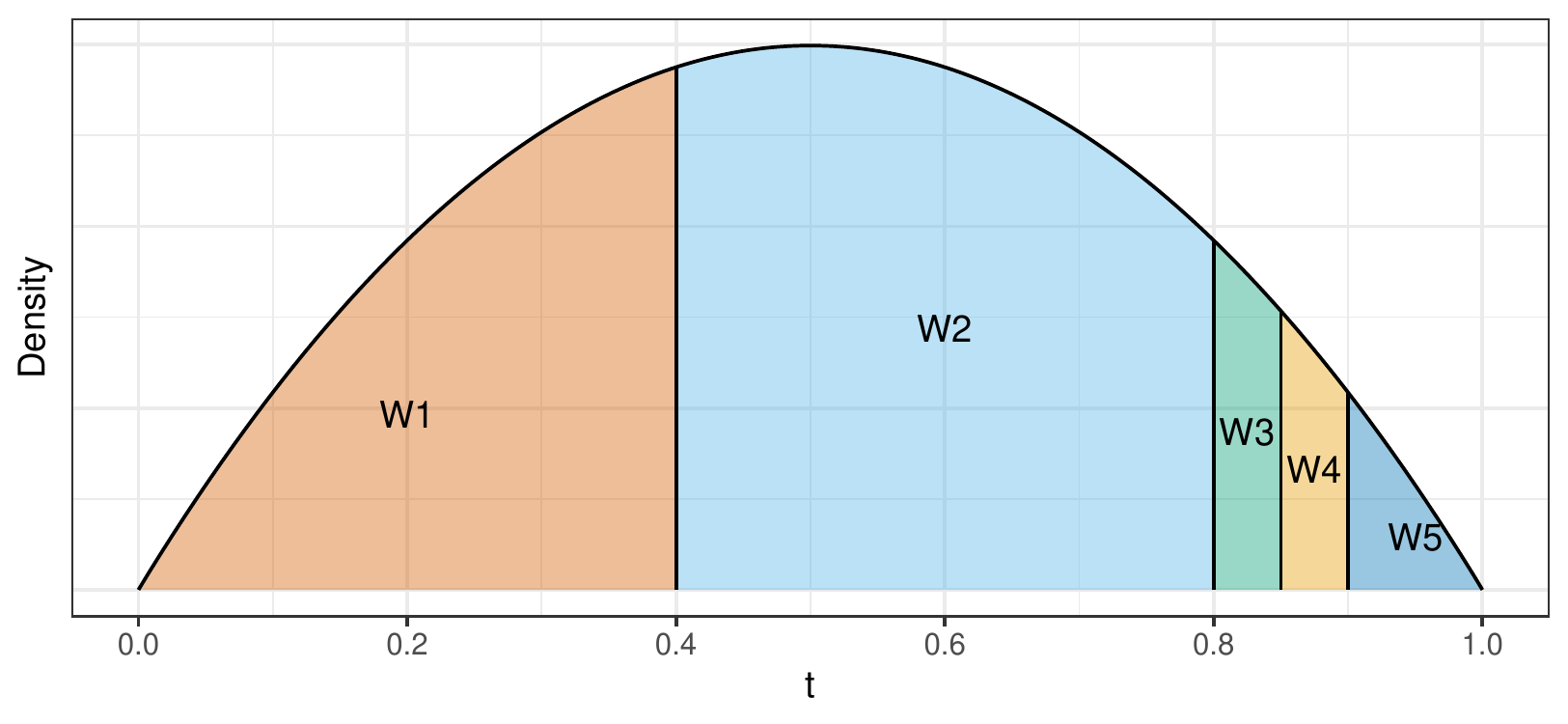} 

}

\caption{Beta(1.993, 1.993) probability density function.}\label{fig:whd2}
\end{figure}

Using \(I_t(1.993, 1.993)\), we can obtain the Harrell--Davis linear coefficients:

\[
W^*_{\operatorname{HD},\{1..5\}} \approx \{ 0.352, 0.543, 0.043, 0.033, 0.028 \}.
\]

Even though we estimate the median,
\(x_{(2)}\) has the highest linear coefficient \(W^*_{\operatorname{HD},2} \approx \{ 0.543 \}\)
due to the high value of \(w_{(2)} = 0.4\).
The second highest linear coefficient \(W^*_{\operatorname{HD},1} \approx \{ 0.352 \}\)
corresponds to the first order statistic \(x_{(1)}\);
it is smaller than \(W^*_{\operatorname{HD},2}\) because it is farther from the center than \(x_{(2)}\),
but it is larger than \(W^*_{\operatorname{HD},3}\) because the interval
\([t^*_0; t^*_1] = [0;0.4]\) is larger and closer to the center than \([t^*_2; t^*_3] = [0.8;0.85]\).

Now we can estimate the median using the linear combination of all order statistics:

\[
\QHD^*(\x, \w, p) = \sum_{i=1}^{n} W^*_{\operatorname{HD},i} \cdot x_{(i)} \approx 1.842.
\]
\end{example}

\clearpage

\hypertarget{sec:wthd}{%
\section{Weighted trimmed Harrell--Davis quantile estimator}\label{sec:wthd}}

The Harrell--Davis quantile estimator is an efficient replacement for traditional quantile estimators
due to its high statistical efficiency (especially for middle quantiles).
However, this estimator is not robust: its breakdown point is zero.
That is why we consider its trimmed modification (see \autocite{akinshin2022thdqe}).
The basic idea is simple: since most of the linear coefficients \(W_{\operatorname{HD},i}\) are pretty small,
they do not have a noticeable impact on efficiency, but they significantly reduce the breakdown point.
In \autocite{akinshin2022thdqe}, it was suggested to build a trimmed modification
based on the beta distribution highest density interval \([L;R]\) of the given size \(D\):

\[
\QTHD(\x, p) = \sum_{i=1}^{n} W_{\THD,i} \cdot x_{(i)}, \quad
W_{\THD,i} = F_{\THD}(t_i) - F_{\THD}(t_{i-1}),\quad
t_i = i / n,
\]

\[
F_{\THD}(t) = \begin{cases}
0 & \textrm{for }\, t < L,\\
\Bigl( I_t(\alpha, \beta) - I_L(\alpha, \beta) \Bigr) / \Bigl( I_R(\alpha, \beta) - I_L(\alpha, \beta) \Bigr)
  & \textrm{for }\, L \leq t \leq R,\\
1 & \textrm{for }\, R < t.
\end{cases}
\]

In the weighted case, we should replace the interval \([L;R]\)
with the highest density interval of \(\Beta(\alpha^*, \beta^*)\),
which we denote by \([L^*;R^*]\).
Adoption of Equation~\eqref{eq:whd} for the trimmed Harrell--Davis quantile estimator is trivial:

\begin{equation}
\QTHD^*(\x, \w, p) = \sum_{i=1}^{n} W^*_{\operatorname{THD},i} \cdot x_{(i)}, \quad
W^*_{\operatorname{THD},i} = F^*_{\operatorname{THD}}(t^*_i) - F^*_{\operatorname{THD}}(t^*_{i-1}),\quad
t^*_i = s_i(\overline{\w}),
\label{eq:wthd}
\end{equation}

\[
F^*_{\operatorname{THD}}(t) = \begin{cases}
0 & \textrm{for }\, t < L^*,\\
\Bigl( I_t(\alpha^*, \beta^*) - I_{L^*}(\alpha^*, \beta^*) \Bigr) / \Bigl( I_{R^*}(\alpha^*, \beta^*) - I_{L^*}(\alpha^*, \beta^*) \Bigr) & \textrm{for }\, L^* \leq t \leq R^*,\\
1 & \textrm{for }\, R^* < t.
\end{cases}
\]

Statistical efficiency and robustness of \(\QTHD\)
are explored in \autocite[Section ``Simulation studies'']{akinshin2022thdqe}.

Note that the trimmed Harrell--Davis quantile estimator has the same limitations as the original estimator:
it is defined only for \(p \in (0;1)\).

The difference between the original Harrell--Davis quantile estimator \(\QHD\)
and its trimmed modification \(\QTHD\) is shown in Example~\ref{exm:wthd}.

\clearpage

\begin{example}
\protect\hypertarget{exm:wthd}{}\label{exm:wthd}\(\x = \{ 1, 2, 3, 10000 \}\), \(\w = \{ 0.1, 0.4, 0.4, 0.1 \}\), \(p = 0.5\).

From~\eqref{eq:kish} and~\eqref{eq:whd-ab}, we get
\(n^* \approx 2.941\), \(\alpha^* = \beta^* \approx 1.971\),
which gives us \(\Beta(1.971, 1.971)\)
(the PDF is shown in Figure~\ref{fig:wthd}).
The cut points are given by:

\[
t^*_{\{0..4\}} = \{ 0, 0.1, 0.5, 0.9, 1 \}.
\]

\begin{figure}[H]

{\centering \includegraphics{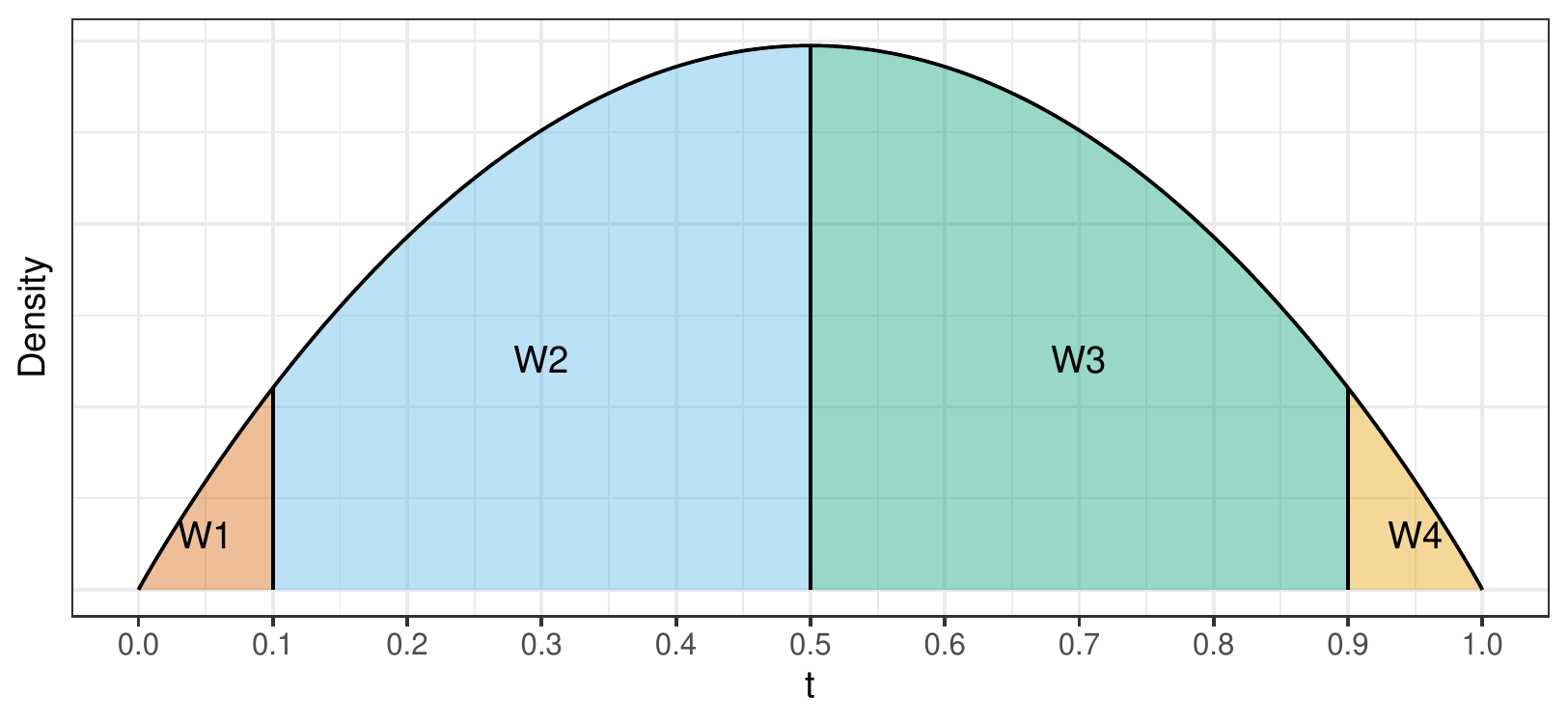} 

}

\caption{Beta(1.971, 1.971) probability density function.}\label{fig:wthd}
\end{figure}

Using \(I_t(1.971, 1.971)\),
we can obtain the linear coefficients for the classic weighted Harrell--Davis quantile estimator:

\[
W^*_{\HD,\{1..4\}} \approx \{ 0.029, 0.471, 0.471, 0.029 \}.
\]

To define \(Q_{\THD}\), we use the rule of thumb from \autocite{akinshin2022thdqe} and set
\(D^* = 1 / \sqrt{n^*} \approx 0.583\).
Since \(\w\) is symmetric, we get \([L^*;R^*] \approx [0.208;0.792]\).
It does not cover \([t^*_0;t^*_1]\) and \([t^*_3;t^*_4]\), therefore \(W^*_{\THD,1} = W^*_{\THD,4} = 0\).
The exact values of the linear coefficients \(W^*_{\THD,i}\) are easy to obtain:

\[
W^*_{\THD,\{1..4\}} = \{ 0, 0.5, 0.5, 0 \}.
\]

Now we can estimate the median using both estimators:

\[
\QHD^*(\x, \w, p) = \sum_{i=1}^{n} W^*_{\operatorname{HD},i} \cdot x_{(i)} \approx 292.594,\quad
\QTHD^*(\x, \w, p) = \sum_{i=1}^{n} W^*_{\operatorname{THD},i} \cdot x_{(i)} = 2.5.
\]

As we can see, the Harrell--Davis median estimation \(\QHD^*(\x, \w, p) \approx 292.594\)
is heavily affected by \(x_{(4)} = 10000\) regardless of its small weight \(w_{(4)} = 0.1\).
The trimmed version of this estimator does not have such a problem thanks to a higher breakdown point:
\(\QTHD^*(\x, \w, p) = 2.5\).
\end{example}

\clearpage

\hypertarget{sec:whf}{%
\section{Weighted traditional quantile estimators}\label{sec:whf}}

Traditionally, statistical packages use various quantile estimation approaches,
that are based on one or two order statistics.
We use the Hyndman--Fan quantile estimator taxonomy (see \autocite{hyndman1996}), which is presented in Table~\ref{tab:hf}.
Here
\(\lfloor\cdot\rfloor\) and \(\lceil\cdot\rceil\) are the floor and ceiling functions;
\(\lfloor\cdot\rceil\) is the banker's rounding
(if the number falls midway between two integers, it is rounded to the nearest even value).
If the obtained \(h\) value is less than~\(1\) or larger than~\(n\), it should be clamped to the \([1;n]\) interval.

\begin{longtable}[]{@{}lll@{}}
\caption{\label{tab:hf} The Hyndman--Fan taxonomy of quantile estimators.}\tabularnewline
\toprule
Type & h & Equation\tabularnewline
\midrule
\endfirsthead
\toprule
Type & h & Equation\tabularnewline
\midrule
\endhead
1 & \(np\) & \(x_{(\hc)}\)\tabularnewline
2 & \(np+1/2\) & \((x_{(\lceil h - 1/2 \rceil)} + x_{(\lceil h + 1/2 \rceil)})/2\)\tabularnewline
3 & \(np\) & \(x_{(\hr)}\)\tabularnewline
4 & \(np\) & \(x_{(\hf)}+(h-\hf)(x_{(\hc)} - x_{(\hf)})\)\tabularnewline
5 & \(np+1/2\) & \(x_{(\hf)}+(h-\hf)(x_{(\hc)} - x_{(\hf)})\)\tabularnewline
6 & \((n+1)p\) & \(x_{(\hf)}+(h-\hf)(x_{(\hc)} - x_{(\hf)})\)\tabularnewline
7 & \((n-1)p+1\) & \(x_{(\hf)}+(h-\hf)(x_{(\hc)} - x_{(\hf)})\)\tabularnewline
8 & \((n+1/3)p+1/3\) & \(x_{(\hf)}+(h-\hf)(x_{(\hc)} - x_{(\hf)})\)\tabularnewline
9 & \((n+1/4)p+3/8\) & \(x_{(\hf)}+(h-\hf)(x_{(\hc)} - x_{(\hf)})\)\tabularnewline
\bottomrule
\end{longtable}

In this classification, only Types 4--7 are continuous;
Types 1--3 have discontinuities, so the corresponding estimators fail to satisfy Requirement~\ref{req:stability}.
Therefore, we consider only Types 4--7 in the rest of the paper (the most popular default option is Type 7).
These estimators use the same interpolation equation:

\begin{equation}
\Q_k(\x, p) = x_{(\hf)}+(h-\hf)(x_{(\hc)} - x_{(\hf)}),
\label{eq:hf1}
\end{equation}

where
\(k\) is the index of the quantile estimator in the Hyndman--Fan taxonomy,
\(h\) is defined in Table~\ref{tab:hf} according to the given \(k\).

\bigskip

We try to adopt the weighted approach proposed in Sections~\ref{sec:whd} and~\ref{sec:wthd}
for \(\QHD\) and \(\QTHD\).
In order to do it, we should express the original non-weighted estimators
in a way similar to the Harrell--Davis quantile estimator.
However, instead of the Beta distribution,
we should define another one that assigns linear coefficients to order statistics.
Since \(\Q_k\) always uses a linear combination of two order statistics,
the most straightforward approach is to use the uniform distribution
\(\mathcal{U}\bigl((h-1)/n, h/n\bigr)\),
which is defined by the following PDF \(f_k\) and CDF \(F_k\):

\[
f_k(t) = \left\{
\begin{array}{lcrcllr}
0      & \textrm{for} &         &      & t  & <    & (h-1)/n, \\
n      & \textrm{for} & (h-1)/n & \leq & t  & \leq & h/n, \\
0      & \textrm{for} & h/n     & <    & t, &      &
\end{array}
\right.
\]

\[
F_k(t) = \left\{
\begin{array}{lcrcllr}
0      & \textrm{for} &         &      & t  & <    & (h-1)/n, \\
tn-h+1 & \textrm{for} & (h-1)/n & \leq & t  & \leq & h/n, \\
1      & \textrm{for} & h/n     & <    & t. &      &
\end{array}
\right.
\]

Now Equation~\eqref{eq:hf1} can be rewritten in a form that matches the definition of \(\QHD\) (Equation~\eqref{eq:hd}):

\begin{equation}
Q_k(\x, p) = \sum_{i=1}^{n} W_{F_k,i} \cdot x_{(i)},\quad
W_{F_k, i} = F_k(t_i) - F_k(t_{i-1}),\quad
t_i = i / n.
\label{eq:hf2}
\end{equation}

Illustrations of Equation~\eqref{eq:hf2} for \(Q_7\) are given in Examples~\ref{exm:hf-a} and~\ref{exm:hf-b}.

\clearpage

\begin{example}[Non-weighted Type 7 quantile estimator, integer h]
\protect\hypertarget{exm:hf-a}{}\label{exm:hf-a}\(n = 5\), \(p = 0.25\).\\
We have integer \(h = 2\), which gives us the PDF and CDF presented in Figure~\ref{fig:hf-a}.

\begin{figure}[H]

{\centering \includegraphics{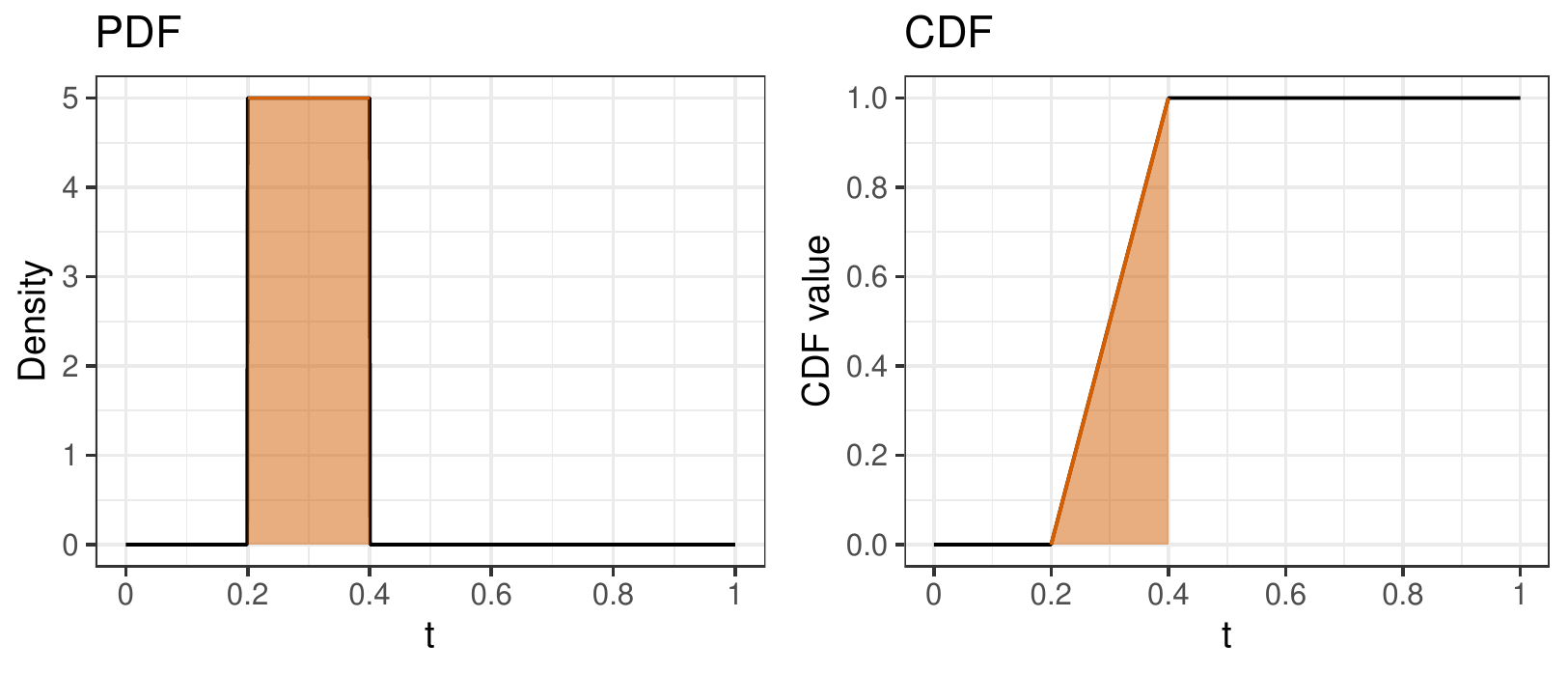} 

}

\caption{$f_7$ and $F_7$ for $n=5$, $p=0.25$.}\label{fig:hf-a}
\end{figure}

It is easy to see that we have only one non-negative \(W_{F_7,i}\): \(W_{F_7,2} = 1\).
This means that \(Q_7(\{ x_1, x_2, x_3, x_4, x_5 \}, 0.25) = x_{(2)}\),
which satisfies our expectations (the first quartile of a sample with five elements is the second element).
\end{example}

\begin{example}[Non-weighted Type 7 quantile estimator, non-integer h]
\protect\hypertarget{exm:hf-b}{}\label{exm:hf-b}\(n = 5\), \(p = 0.35\).\\
We have non-integer \(h = 2.4\), which gives us the PDF and CDF presented in Figure~\ref{fig:hf-b}.

\begin{figure}[H]

{\centering \includegraphics{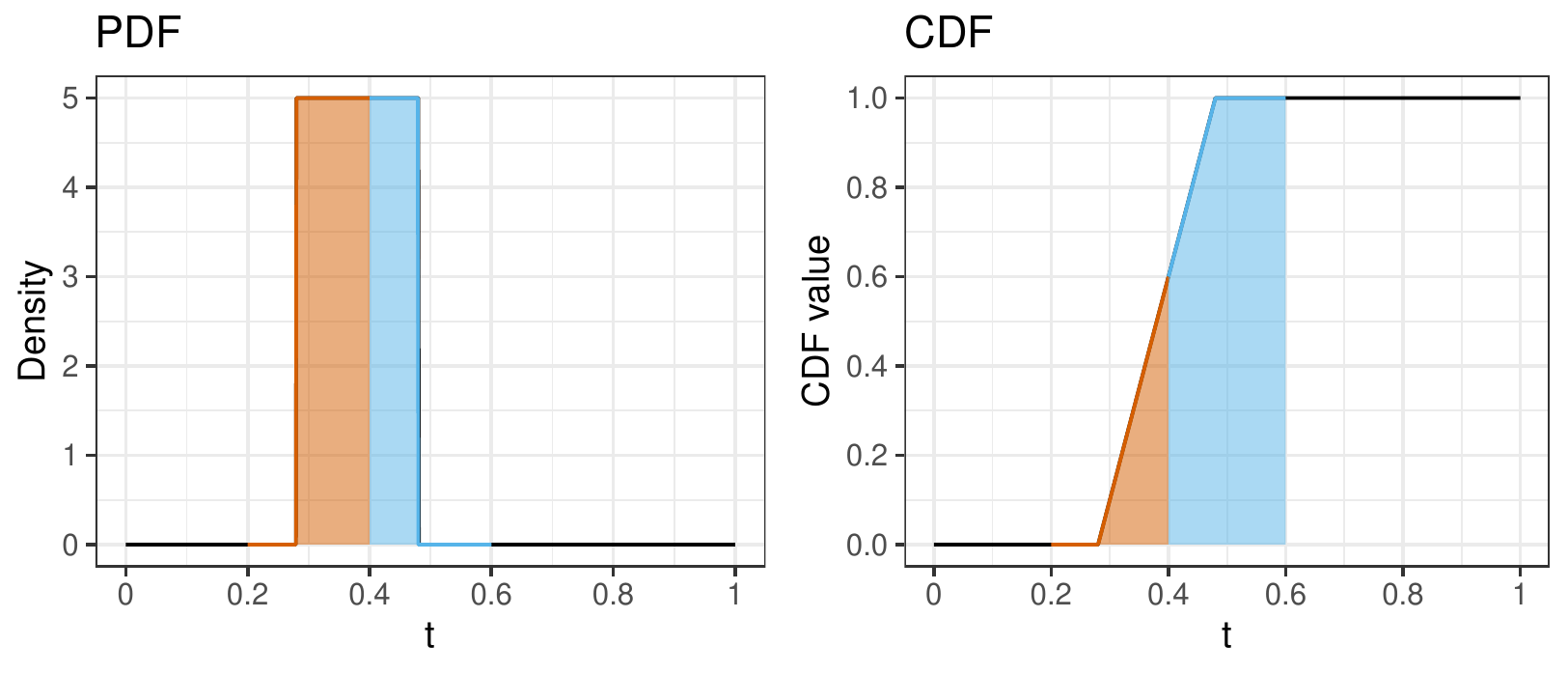} 

}

\caption{$f_7$ and $F_7$ for $n=5$, $p=0.35$.}\label{fig:hf-b}
\end{figure}

As we can see, the \(0.35^\textrm{th}\) quantile estimation is a linear combination of \(x_{(2)}\) and \(x_{(3)}\)
with coefficients \(W_{F_7, 2} = 0.6\), \(W_{F_7, 3} = 0.4\).
Thus, \(Q_7(\{ x_1, x_2, x_3, x_4, x_5 \}, 0.35) = 0.6 \cdot x_{(2)} + 0.4 \cdot x_{(3)}\).
\end{example}

\clearpage

As we can see, the suggested \(F_k\) perfectly matches the non-weighted case.
Now we can generalize Equation~\eqref{eq:hf2} to the weighted case
similarly to Equations~\eqref{eq:whd} and~\eqref{eq:wthd}
(assuming \(h^*\) is obtained from Table~\ref{tab:hf} based on \(n^*\)):

\[
Q_k^*(\x, \w, p) = \sum_{i=1}^{n} W^*_{F^*_k,i} \cdot x_{(i)},\quad
W^*_{F^*_k, i} = F^*_k(t^*_i) - F^*_k(t^*_{i-1}),\quad
t^*_i = s_i(\overline{\w}),
\]

\[
F^*_k(t) = \left\{
\begin{array}{lcrcllr}
0          & \textrm{for} &             &      & t  & <    & (h^*-1)/n^*, \\
tn^*-h^*+1 & \textrm{for} & (h^*-1)/n^* & \leq & t  & \leq & h^*/n^*, \\
1          & \textrm{for} & h^*/n^*     & <    & t. &      &
\end{array}
\right.
\]

The usage of \(Q_7^*(\x, \w, p)\) is shown in Example~\ref{exm:hf-c}.
More examples can be found in Section~\ref{sec:sim}.

\begin{example}[Weighted Type 7 quantile estimator]
\protect\hypertarget{exm:hf-c}{}\label{exm:hf-c}

\({\x = \{ 1, 2, 3, 4, 5\},\, \w = \{ 0.3, 0.1, 0, 0.1, 0.4 \},\, p = 0.5}\).\\
In this case, \(n^* = 3\), \(h^* = (n^*-1)p + 1 = 2\), which gives us the following expression for \(F^*_7\):

\[
F^*_7(t) = \left\{
\begin{array}{lcrcllr}
0    & \textrm{for} &     &      & u  & <    & 1/3, \\
3t-1 & \textrm{for} & 1/3 & \leq & u  & \leq & 2/3, \\
1    & \textrm{for} & 2/3 & <    & u. &      &
\end{array}
\right.
\]

Now we can calculate the values of the intermediate variables:

\[
\begin{aligned}
\overline{\w}         & = \{ 3/9,\; 1/9,\; 0,\; 1/9,\; 4/9 \},\\
t^*_{\{0..5\}}        & = \{ 0,\; 3/9,\; 4/9,\; 4/9,\; 5/9,\; 1 \},\\
F^*_7(t^*_{\{0..5\}}) & = \{ 0,\; 0,\; 1/3,\; 1/3,\; 2/3,\; 1 \},\\
W^*_{F^*_7, \{1..5\}} & = \{ 0,\; 1/3,\; 0,\; 1/3,\; 1/3 \}.
\end{aligned}
\]

Finally, the median estimation is defined as follows:

\[
Q_7^*(\x, \w, p) =
  \sum_{i=1}^n W^*_{F^*_7, i} \cdot x_{(i)} =
  (x_{(2)} + x_{(4)} + x_{(5)}) / 3 =
  11/3 \approx
  3.666667.
\]

This example is illustrated in Figure~\ref{fig:hf-c}.

\begin{figure}[H]

{\centering \includegraphics{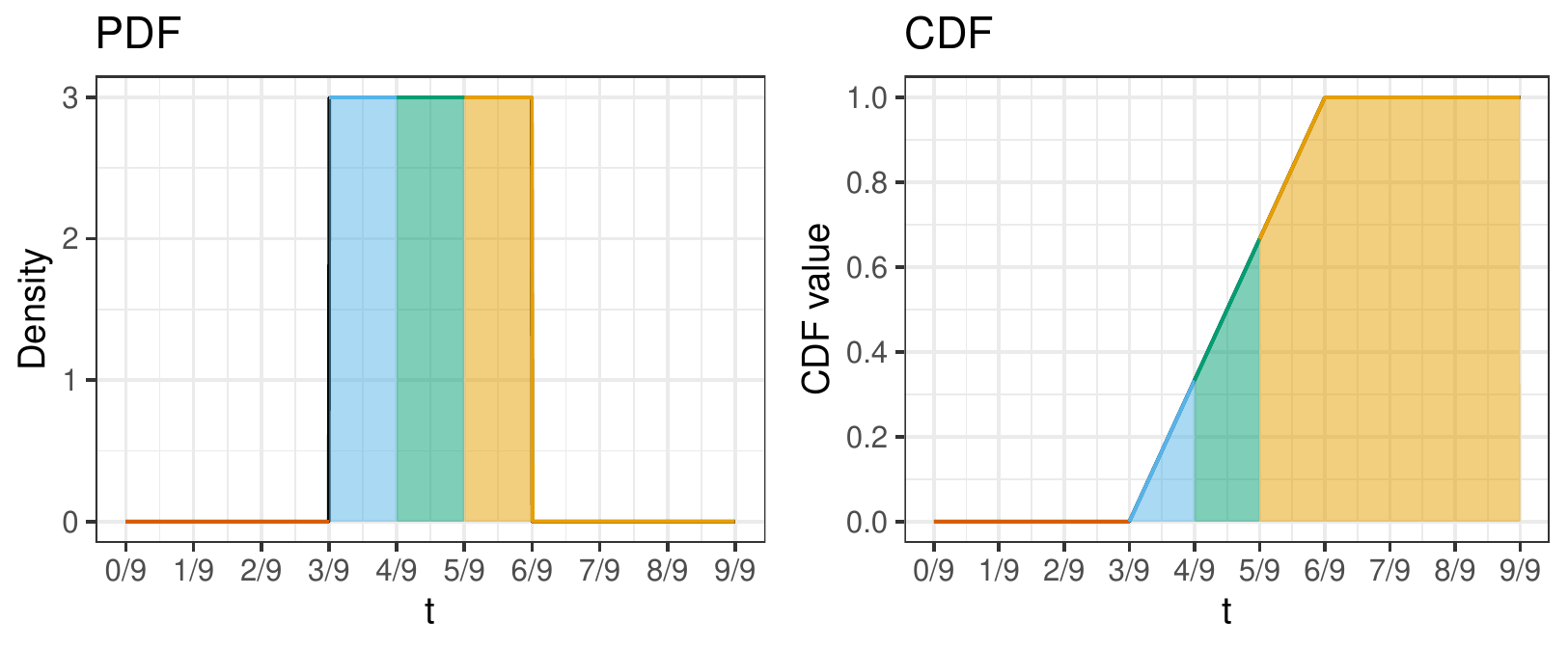} 

}

\caption{$f^*_7$ and $F^*_7$ for $\w=\{ 0.3, 0.1, 0, 0.1, 0.4 \}$, $p=0.5$.}\label{fig:hf-c}
\end{figure}

\end{example}

\clearpage

\hypertarget{sec:sim}{%
\section{Simulation studies}\label{sec:sim}}

In this section, we perform several simulation studies
in order to show the practical usage examples of the suggested approach in various contexts.
These simulations are not comprehensive research of the estimator properties
but rather just an illustration demonstrating practical applications of the weighted quantile estimators.

\hypertarget{sec:sim-mes}{%
\subsection{Simulation 1: Median exponential smoothing}\label{sec:sim-mes}}

In this study, we discuss the problem of quantile exponential smoothing
and consider \(\x\) as a time series of some measurements
(the oldest measurements are at the beginning of the series, and the newest measurements are at the end).
In such situations, it is convenient to assign weights according to the exponential decay law:

\begin{equation}
\omega(t) = 2^{-t/t_{1/2}},
\label{eq:decay-exp}
\end{equation}

where the parameter \(t_{1/2}\) is known as the half-life.
This value describes the period required for the current weight to reduce to half of its original value.
Thus, we have

\[
\omega(0) = 1, \quad \omega(t_{1/2}) = 0.5, \quad \omega(2t_{1/2}) = 0.25, \quad \omega(3t_{1/2}) = 0.125, \quad \ldots
\]

The simplest way to define \(\w\) for our time series is to apply the exponential decay law in a reverse way:

\begin{equation}
w_i = \omega(n - i).
\label{eq:decay-rev}
\end{equation}

\begin{figure}[H]

{\centering \includegraphics{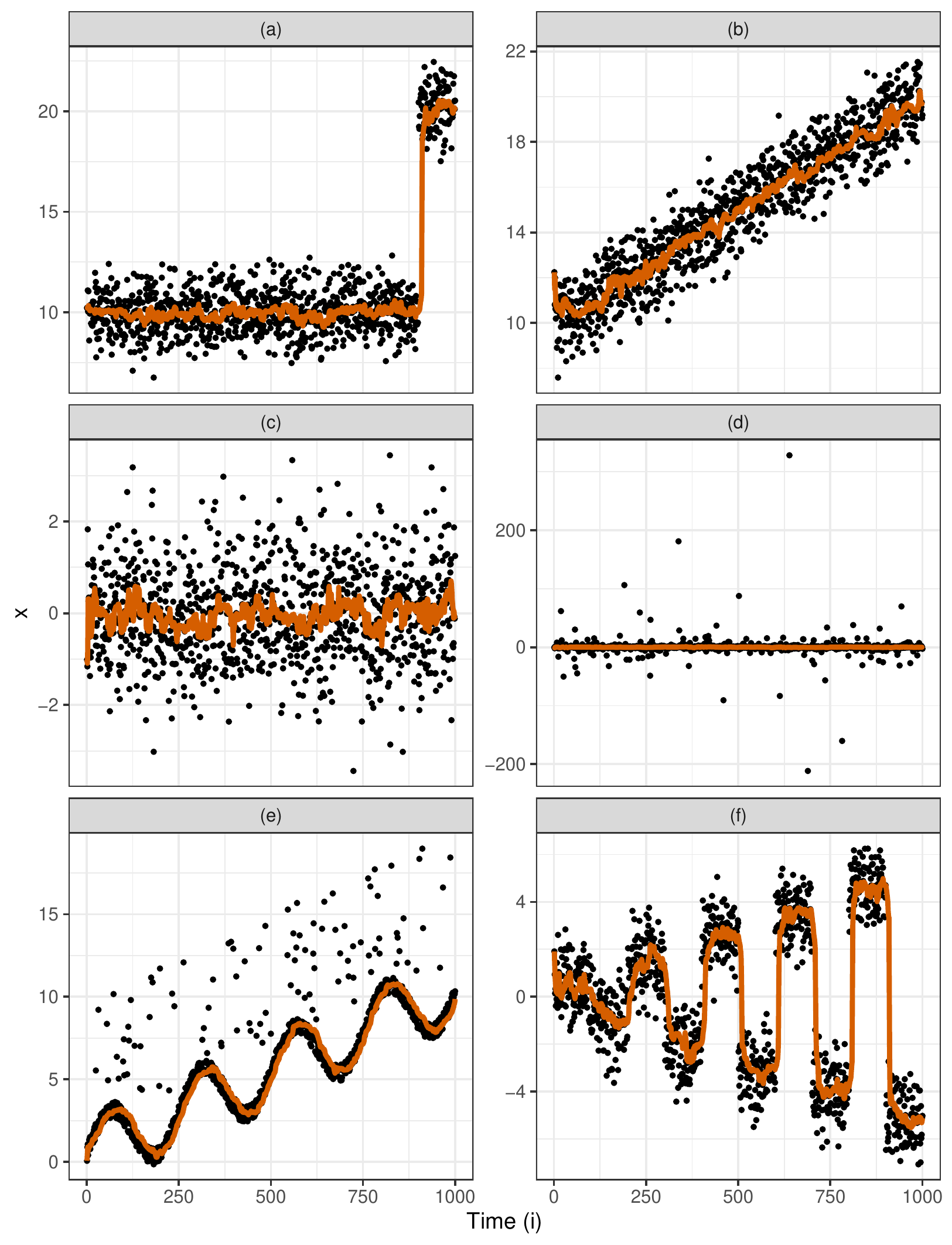} 

}

\caption{Median exponential smoothing.}\label{fig:simf-mes}
\end{figure}

An example of the median exponential smoothing for various time series
is presented in Figure~\ref{fig:simf-mes}
(\(n = 1\,000\), \(t_{1/2}=10\), the weights are assigned according to Equation~\eqref{eq:decay-rev}).
The following time series are considered:

\begin{itemize}
\item
  \begin{enumerate}
  \def\labelenumi{(\alph{enumi})}
  \tightlist
  \item
    Two normal distributions split by a change point:
    \(x_i \in \mathcal{N}(10, 1)\) for \(1 \leq i \leq 900\); \(x_i \in \mathcal{N}(20, 1)\) for \(901 \leq i \leq 1\,000\);
  \end{enumerate}
\item
  \begin{enumerate}
  \def\labelenumi{(\alph{enumi})}
  \setcounter{enumi}{1}
  \tightlist
  \item
    A moving normal distribution: \(x_i \in \mathcal{N}(10 + i / 100, 1)\);
  \end{enumerate}
\item
  \begin{enumerate}
  \def\labelenumi{(\alph{enumi})}
  \setcounter{enumi}{2}
  \tightlist
  \item
    The standard normal distribution: \(x_i \in \mathcal{N}(0, 1)\);
  \end{enumerate}
\item
  \begin{enumerate}
  \def\labelenumi{(\alph{enumi})}
  \setcounter{enumi}{3}
  \tightlist
  \item
    The standard Cauchy distribution: \(x_i \in \operatorname{Cauchy}(0, 1)\);
  \end{enumerate}
\item
  \begin{enumerate}
  \def\labelenumi{(\alph{enumi})}
  \setcounter{enumi}{4}
  \tightlist
  \item
    A monotonically increasing sine wave pattern with occasional outliers;
  \end{enumerate}
\item
  \begin{enumerate}
  \def\labelenumi{(\alph{enumi})}
  \setcounter{enumi}{5}
  \tightlist
  \item
    ``Dispersing'' normal distributions:
    \(x_i \in \mathcal{N}\Bigl(\bigl( 2\cdot (j \bmod 2) - 1 \bigr) \cdot \bigl(j / 2 + ((i - 1) \bmod 100) / 100 \bigr) , 1\Bigr)\),
    where \(j = \lfloor (i - 1) / 100 \rfloor + 1\).
  \end{enumerate}
\end{itemize}

For each time series, we consider all the subseries that start at the beginning: \(\{ x_1, x_2, \ldots, x_i\}\).
For each subseries, we estimate the median using \(Q_7^*\).
Thus, we get the moving (running) median shown in Figure~\ref{fig:simf-mes} with a bold line.
As we can see, the obtained estimations match our expectations:
the moving median follows the trends and quickly recovers after change points.

In practical applications, more sophisticated weight assignation strategies can be used.
For example, if we collect a group of measurements daily,
and we know that each group is obtained from the same distribution
(no changes can be introduced between measurements within the same group),
the exponential law can be applied on a per-group basis
(all the group elements get the same weight).

\clearpage

\clearpage

\hypertarget{sec:sim-qes}{%
\subsection{Simulation 2: Quartile exponential smoothing}\label{sec:sim-qes}}

In this simulation, we consider a time series consisting of four parts:

\begin{itemize}
\tightlist
\item
  \((S_1)\; i \in \{ \phantom{00}1..100 \}:\; x_i \in \mathcal{N}(0, 1)\);
\item
  \((S_2)\; i \in \{ 101..200 \}:\; x_i \in \mathcal{N}(10, 1)\);
\item
  \((S_3)\; i \in \{ 201..300 \}:\; x_i \in \mathcal{N}(0, 1)\);
\item
  \((S_4)\; i \in \{ 301..500 \}:\;  x_i \in 0.4\cdot\mathcal{N}(-10, 1) + 0.2\cdot\mathcal{N}(0, 1) + 0.4\cdot\mathcal{N}(10, 1)\).
\end{itemize}

Similar to the previous simulation, we calculate running estimations.
In addition to the running median (which is also the second quartile and the \(50^\textrm{th}\) percentile),
we consider the first and the third running quartiles
(which are also the \(25^\textrm{th}\) and \(75^\textrm{th}\) percentiles respectively).
The weights \(w_i\) are assigned according to the exponential law
defined in Equations~\eqref{eq:decay-exp} and~\eqref{eq:decay-rev}.
We enumerate three half-life values: (a) \(t_{1/2}=5\), (b) \(t_{1/2}=10\), (c) \(t_{1/2}=30\).

\begin{figure}[H]

{\centering \includegraphics{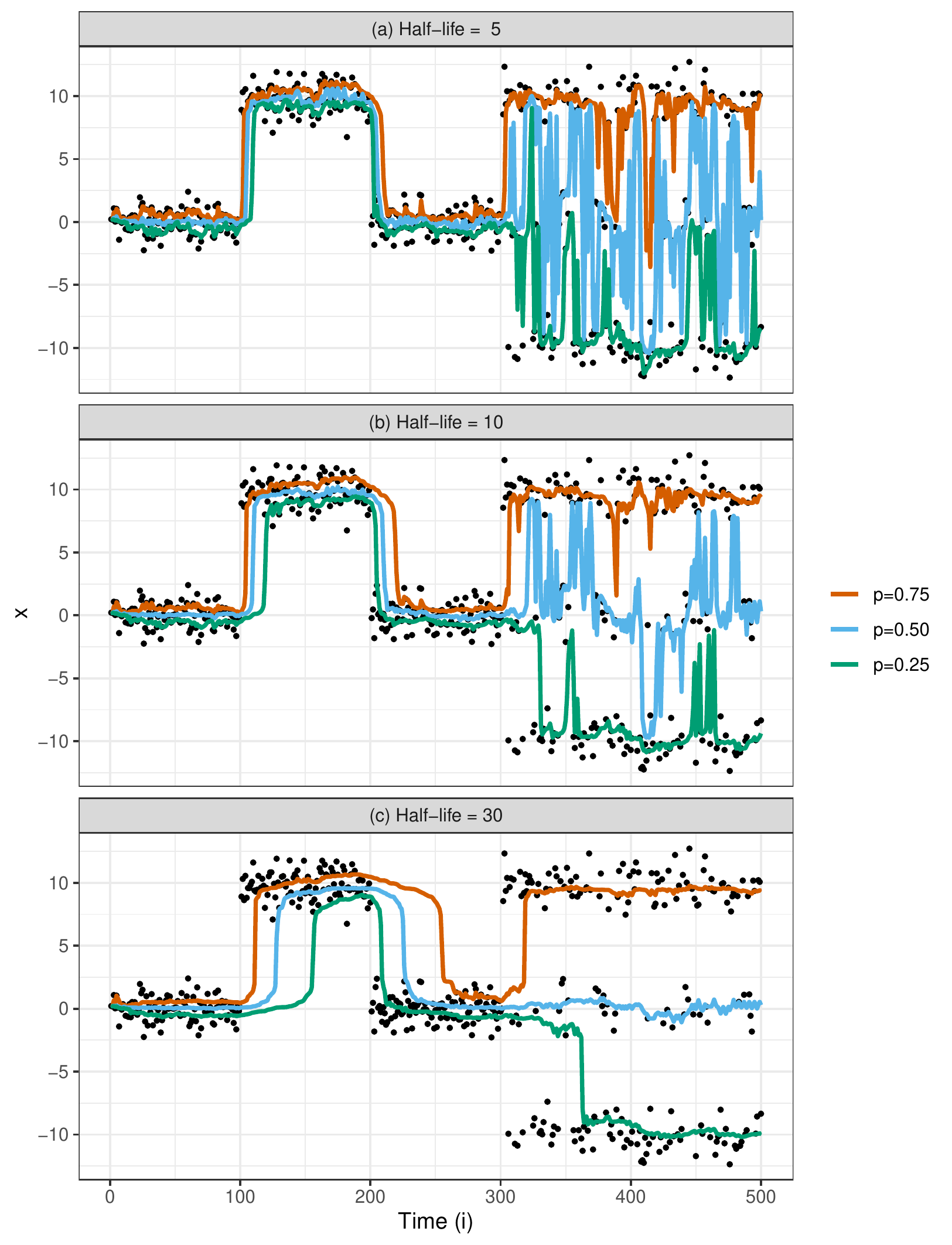} 

}

\caption{Quartile exponential smoothing.}\label{fig:simf-qes}
\end{figure}

The simulation results are presented in Figure~\ref{fig:simf-qes}.
Based on these plots, we can make the following observations:

\begin{itemize}
\tightlist
\item
  In the context of quantile exponential smoothing, the running median is the most reliable metric:
  it quickly adapts to various distribution changes.
\item
  Higher and lower running quantiles perform worse:
  the adaptation period to the distribution changes heavily depends on the direction of these changes.
\item
  Small half-life values shorten the adaptation period but increase the dispersion of estimations.
  Large half-life values enlarge the adaptation period but decrease the dispersion of estimations.
  If the target quantile is around a low-density region, it needs a larger half-life value to avoid jumping
  between neighboring high-density regions.
\end{itemize}

\clearpage

\clearpage

\hypertarget{sec:sim-mix}{%
\subsection{Simulation 3: Weighted mixture distribution}\label{sec:sim-mix}}

As we have discussed in Section~\ref{sec:pre},
the problems of quantile exponential smoothing and the weighted mixture distribution quantile estimation
may have different expectations of the quantile values (see Example~\ref{exm:smoothing-vs-mixture}).
However, the issues typically arise when we estimate a single quantile value in a low-density region.
If we are estimating all values of the quantile function or
if we are interested in quantile values in high-density regions,
the described weighted quantile estimators usually produce reliable results.
Let us conduct one more simulation study to illustrate these cases.

In this simulation, we consider the following six mixture distributions:

\begin{itemize}
\item
  \begin{enumerate}
  \def\labelenumi{(\alph{enumi})}
  \tightlist
  \item
    A mixture of two normal distributions with a small gap:
    \(0.75 \cdot \mathcal{N}(0, 1) + 0.25 * \mathcal{N}(5, 3)\);
  \end{enumerate}
\item
  \begin{enumerate}
  \def\labelenumi{(\alph{enumi})}
  \setcounter{enumi}{1}
  \tightlist
  \item
    A mixture of two normal distributions with a large gap:
    \(0.99 \cdot \mathcal{N}(0, 1) + 0.01 \cdot \mathcal{N}(100, 10)\);
  \end{enumerate}
\item
  \begin{enumerate}
  \def\labelenumi{(\alph{enumi})}
  \setcounter{enumi}{2}
  \tightlist
  \item
    A mixture of two uniform distributions with a small gap:
    \(0.5 \cdot \mathcal{U}(0, 1) + 0.5 \cdot \mathcal{U}(5, 10)\);
  \end{enumerate}
\item
  \begin{enumerate}
  \def\labelenumi{(\alph{enumi})}
  \setcounter{enumi}{3}
  \tightlist
  \item
    A mixture of two uniform distributions with a large gap:
    \(0.1 \cdot \mathcal{U}(0, 1) + 0.9 \cdot \mathcal{U}(20, 30)\);
  \end{enumerate}
\item
  \begin{enumerate}
  \def\labelenumi{(\alph{enumi})}
  \setcounter{enumi}{4}
  \tightlist
  \item
    A mixture of three exponential distributions:
    \(0.7 \cdot \Exp(\lambda = 1) + 0.2 \cdot \Exp(\lambda = 2) + 0.1 \cdot \Exp(\lambda = 3)\);
  \end{enumerate}
\item
  \begin{enumerate}
  \def\labelenumi{(\alph{enumi})}
  \setcounter{enumi}{5}
  \tightlist
  \item
    A mixture of three shifted exponential distributions:\\
    \(0.3 \cdot \Exp(\lambda = 1,\, \textrm{shift} = 0) +  0.3 \cdot \Exp(\lambda = 1,\, \textrm{shift} = 10) +  0.4 \cdot \Exp(\lambda = 1,\, \textrm{shift} = 20)\).
  \end{enumerate}
\end{itemize}

For each mixture, we generate random samples of size \(100\)
from each individual distribution that contributes to the mixture.
Next, we combine them into a weighted sample according to the individual distribution weights.
For example, for the mixture (a), we get a sample of size \(200\), in which
\(x_{\{1..100\}} \in \mathcal{N}(0, 1)\),
\(x_{\{101..200\}} \in \mathcal{N}(5, 3)\),
\(w_{\{1..100\}} = 0.75\),
\(w_{\{101..200\}} = 0.25\).
Finally, we estimate the mixture quantiles using the weighted Hyndman--Fan Type 7 estimator for
\(p \in [0.01, 0.99]\) with step \(\Delta p = 0.01\).
We perform \(50\) independent trials according to the described procedure.
Thus, we get \(50\) quantile function estimations per mixture.

\begin{figure}[H]

{\centering \includegraphics{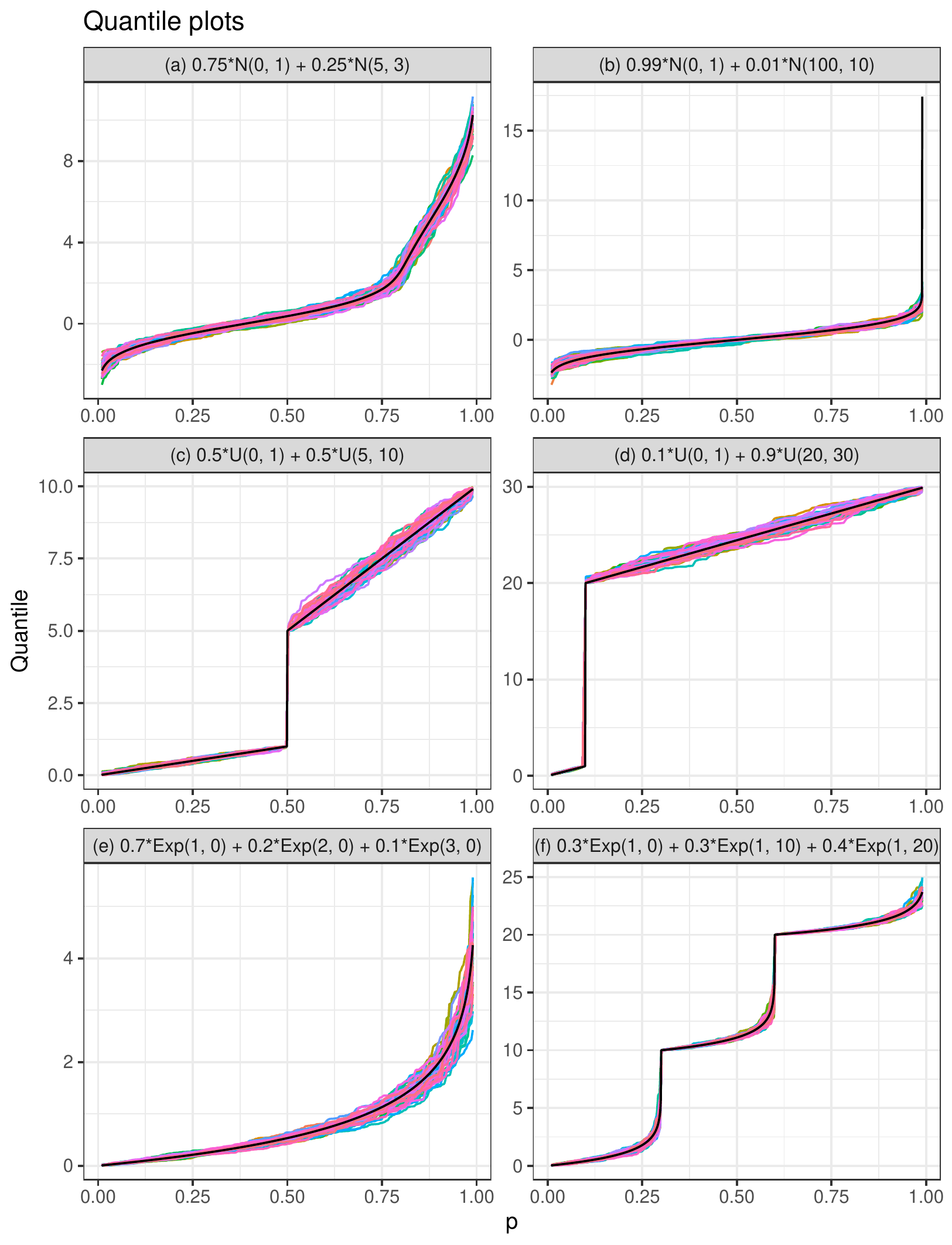} 

}

\caption{Quantile plots of weighted mixture distributions.}\label{fig:simf-mix1}
\end{figure}

In Figure~\ref{fig:simf-mix1}, we can see the obtained quantile functions
(the true quantiles are shown in black, the estimated quantiles are shown in color).
As we can see, the estimations are quite accurate.
There is some divergence from the true quantiles due to randomness and the small sample sizes,
but the distribution form is still recognizable.

\begin{figure}[H]

{\centering \includegraphics{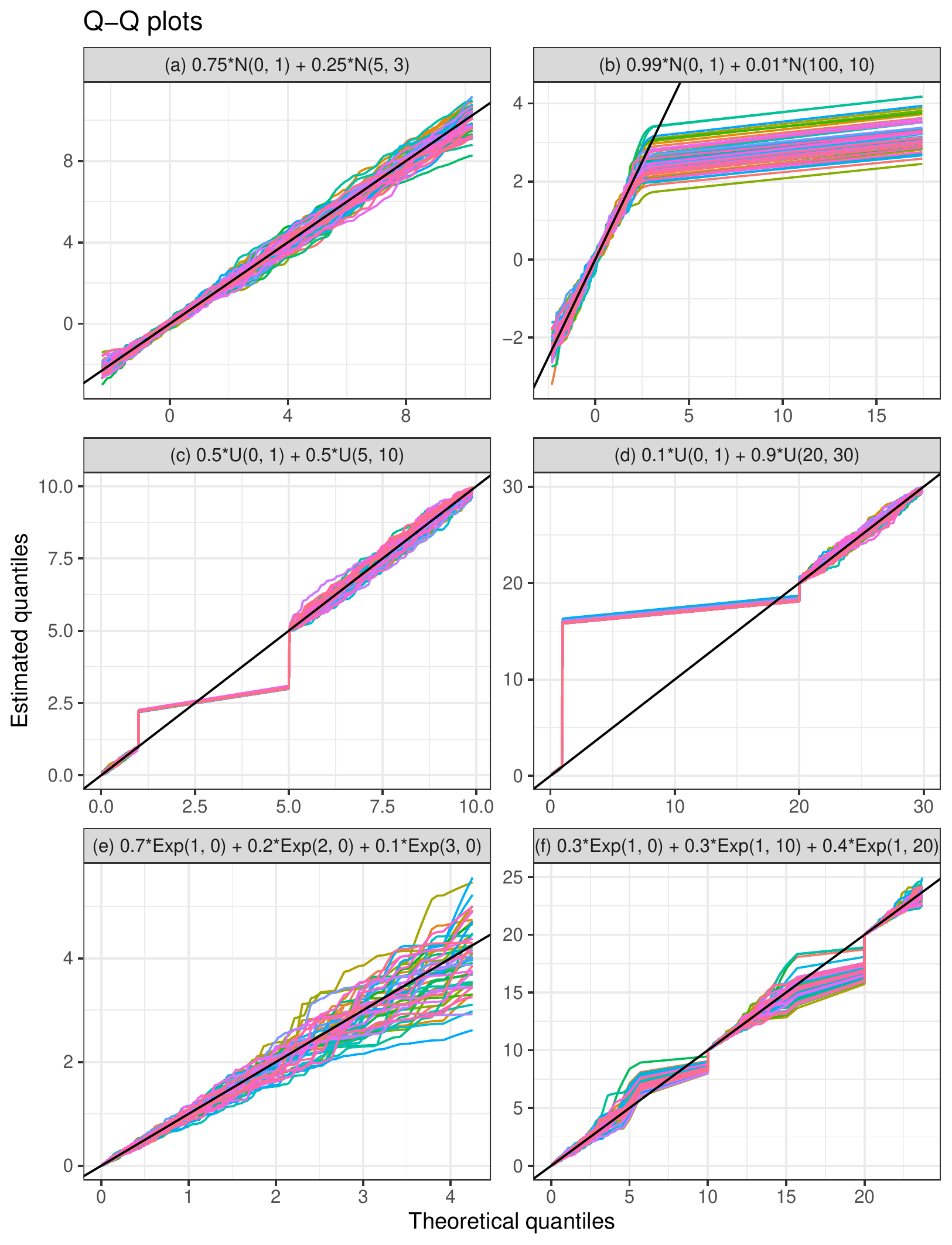} 

}

\caption{Q-Q plots of weighted mixture distributions.}\label{fig:simf-mix2}
\end{figure}

\begin{figure}[H]

{\centering \includegraphics{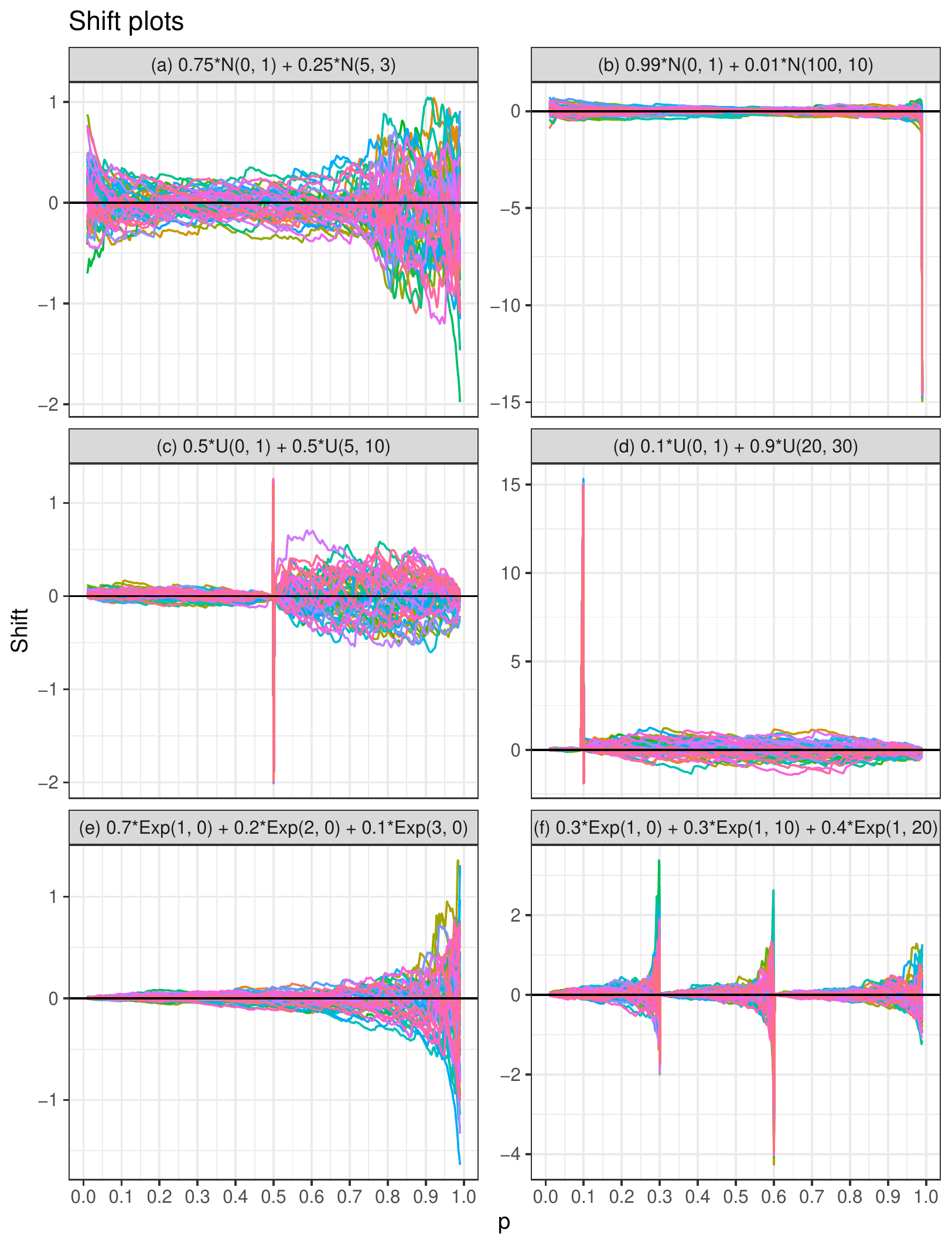} 

}

\caption{Shift plots of weighted mixture distributions.}\label{fig:simf-mix3}
\end{figure}

However, if we are interested not in the form of the whole distribution
but rather in the values of individual quantiles,
we should look at
the Q-Q plot (see \autocite{chambers1983}) presented in Figure~\ref{fig:simf-mix2}
(the identity line is shown in black, the quantile lines are shown in color)
and the Doksum's shift plots (see \autocite{doksum1974}) presented in Figure~\ref{fig:simf-mix3}
(the zero shift is shown in black, the estimated quantile shifts are shown in color).
As we can see, the quantile estimations near the low-density regions in the gaps between individual distributions
are often not accurate: they significantly diverge from the true quantile values.
It is a common issue not only for weighted quantile estimators but for any kind of quantile estimators.
If we are aware of the fact that the target quantile is near a low-density region
(e.g., it falls within a gap between individual distributions),
we should probably use another estimating procedure
and take the huge dispersion of estimations into account.

In this particular simulation, we have used large sample sizes,
but the suggested weighted quantile estimator can be applied to samples of any size, including \(n=1\)
(unlike the straightforward method described in Section~\ref{sec:pre}).

\clearpage

\clearpage

\clearpage

\clearpage

\hypertarget{sec:summary}{%
\section{Conclusion}\label{sec:summary}}

In this paper, we have discussed an approach for building weighted versions of existing quantile estimators.
In particular, we have considered
the traditional Hyndman--Fan Types 4--7 quantile estimators (extremely robust, not so efficient),
the Harrell--Davis quantile estimator (efficient but not robust),
and the trimmed Harrell--Davis quantile estimator (allows customizing trade-off between robustness and efficiency).
However, the proposed scheme is generic and can be applied to other quantile estimators
based on a linear combination of order statistics.

The suggested approach has several useful properties:
it is a consistent extension of the existing quantile estimators (Requirement~\ref{req:consistency}),
sample elements with zero weights have no impact on the estimation (Requirement~\ref{req:zero}),
small changes in weight coefficients produce small changes in the estimation (Requirement~\ref{req:stability}).
These properties make the proposed estimators particularly beneficial for quantile exponential smoothing,
but they can also be applied to other problems,
such as estimating quantiles of a mixture distribution based on individual samples.

The obtained weighted quantile estimations can be used as a base for other estimators, such as
various location estimators (e.g.,
Tukey's trimean (\autocite{tukey1977}),
Midsummary (\autocite{tukey1977}),
Midhinge (\autocite{tukey1977}),
Gastwirth's location estimator (\autocite{gastwirth1966})) and
scale estimators (e.g.,
interquartile and interdecile ranges,
median absolute deviation (\autocite{hampel1974,rousseeuw1993,wilcox2016,akinshin2022madfactors}),
quantile absolute deviation (\autocite{akinshin2022qad})).
The exponential smoothing can be easily applied to any of the derived estimators.

\hypertarget{disclosure-statement}{%
\section*{Disclosure statement}\label{disclosure-statement}}
\addcontentsline{toc}{section}{Disclosure statement}

The author declares no conflict of interest.

\hypertarget{data-and-source-code-availability}{%
\section*{Data and source code availability}\label{data-and-source-code-availability}}
\addcontentsline{toc}{section}{Data and source code availability}

The source code of this paper and all simulations are available on GitHub:\\
\url{https://github.com/AndreyAkinshin/paper-wqe}.

\hypertarget{acknowledgments}{%
\section*{Acknowledgments}\label{acknowledgments}}
\addcontentsline{toc}{section}{Acknowledgments}

The author thanks Ivan Pashchenko for valuable discussions.

\clearpage

\hypertarget{appendix-appendix}{%
\appendix}

\hypertarget{sec:refimpl}{%
\section{Reference implementation}\label{sec:refimpl}}

The following is an R implementation of all the proposed weighted quantile estimators:

\begin{Shaded}
\begin{Highlighting}[]
\CommentTok{\# Kish\textquotesingle{}s effective sample size}
\NormalTok{kish\_ess \textless{}{-}}\StringTok{ }\ControlFlowTok{function}\NormalTok{(w) }\KeywordTok{sum}\NormalTok{(w)}\OperatorTok{\^{}}\DecValTok{2} \OperatorTok{/}\StringTok{ }\KeywordTok{sum}\NormalTok{(w}\OperatorTok{\^{}}\DecValTok{2}\NormalTok{)}

\CommentTok{\# Weighted generic quantile estimator}
\NormalTok{wquantile\_generic \textless{}{-}}\StringTok{ }\ControlFlowTok{function}\NormalTok{(x, w, probs, cdf) \{}
\NormalTok{  n \textless{}{-}}\StringTok{ }\KeywordTok{length}\NormalTok{(x)}
  \ControlFlowTok{if}\NormalTok{ (}\KeywordTok{is.null}\NormalTok{(w)) \{}
\NormalTok{    w \textless{}{-}}\StringTok{ }\KeywordTok{rep}\NormalTok{(}\DecValTok{1} \OperatorTok{/}\StringTok{ }\NormalTok{n, n)}
\NormalTok{  \}}
  \ControlFlowTok{if}\NormalTok{ (}\KeywordTok{any}\NormalTok{(}\KeywordTok{is.na}\NormalTok{(x))) \{}
\NormalTok{    w \textless{}{-}}\StringTok{ }\NormalTok{w[}\OperatorTok{!}\KeywordTok{is.na}\NormalTok{(x)]}
\NormalTok{    x \textless{}{-}}\StringTok{ }\NormalTok{x[}\OperatorTok{!}\KeywordTok{is.na}\NormalTok{(x)]}
\NormalTok{  \}}

\NormalTok{  nw \textless{}{-}}\StringTok{ }\KeywordTok{kish\_ess}\NormalTok{(w)}

\NormalTok{  indexes \textless{}{-}}\StringTok{ }\KeywordTok{order}\NormalTok{(x)}
\NormalTok{  x \textless{}{-}}\StringTok{ }\NormalTok{x[indexes]}
\NormalTok{  w \textless{}{-}}\StringTok{ }\NormalTok{w[indexes]}

\NormalTok{  w \textless{}{-}}\StringTok{ }\NormalTok{w }\OperatorTok{/}\StringTok{ }\KeywordTok{sum}\NormalTok{(w)}
\NormalTok{  t \textless{}{-}}\StringTok{ }\KeywordTok{cumsum}\NormalTok{(}\KeywordTok{c}\NormalTok{(}\DecValTok{0}\NormalTok{, w))}

  \KeywordTok{sapply}\NormalTok{(probs, }\ControlFlowTok{function}\NormalTok{(p) \{}
\NormalTok{    cdf\_values \textless{}{-}}\StringTok{ }\KeywordTok{cdf}\NormalTok{(nw, p, t)}
\NormalTok{    W \textless{}{-}}\StringTok{ }\KeywordTok{tail}\NormalTok{(cdf\_values, }\DecValTok{{-}1}\NormalTok{) }\OperatorTok{{-}}\StringTok{ }\KeywordTok{head}\NormalTok{(cdf\_values, }\DecValTok{{-}1}\NormalTok{)}
    \KeywordTok{sum}\NormalTok{(W }\OperatorTok{*}\StringTok{ }\NormalTok{x)}
\NormalTok{  \})}
\NormalTok{\}}

\CommentTok{\# Weighted Harrell{-}Davis quantile estimator}
\NormalTok{whdquantile \textless{}{-}}\StringTok{ }\ControlFlowTok{function}\NormalTok{(x, w, probs) \{}
\NormalTok{  cdf \textless{}{-}}\StringTok{ }\ControlFlowTok{function}\NormalTok{(n, p, t) \{}
    \ControlFlowTok{if}\NormalTok{ (p }\OperatorTok{==}\StringTok{ }\DecValTok{0} \OperatorTok{||}\StringTok{ }\NormalTok{p }\OperatorTok{==}\StringTok{ }\DecValTok{1}\NormalTok{)}
      \KeywordTok{return}\NormalTok{(}\KeywordTok{rep}\NormalTok{(}\OtherTok{NA}\NormalTok{, }\KeywordTok{length}\NormalTok{(t)))}
    \KeywordTok{pbeta}\NormalTok{(t, (n }\OperatorTok{+}\StringTok{ }\DecValTok{1}\NormalTok{) }\OperatorTok{*}\StringTok{ }\NormalTok{p, (n }\OperatorTok{+}\StringTok{ }\DecValTok{1}\NormalTok{) }\OperatorTok{*}\StringTok{ }\NormalTok{(}\DecValTok{1} \OperatorTok{{-}}\StringTok{ }\NormalTok{p))}
\NormalTok{  \}}
  \KeywordTok{wquantile\_generic}\NormalTok{(x, w, probs, cdf)}
\NormalTok{\}}

\CommentTok{\# Weighted trimmed Harrell{-}Davis quantile estimator}
\NormalTok{wthdquantile \textless{}{-}}\StringTok{ }\ControlFlowTok{function}\NormalTok{(x, w, probs, }\DataTypeTok{width =} \DecValTok{1} \OperatorTok{/}\StringTok{ }\KeywordTok{sqrt}\NormalTok{(}\KeywordTok{kish\_ess}\NormalTok{(w)))}
                  \KeywordTok{sapply}\NormalTok{(probs, }\ControlFlowTok{function}\NormalTok{(p) \{}
\NormalTok{  getBetaHdi \textless{}{-}}\StringTok{ }\ControlFlowTok{function}\NormalTok{(a, b, width) \{}
\NormalTok{    eps \textless{}{-}}\StringTok{ }\FloatTok{1e{-}9}
    \ControlFlowTok{if}\NormalTok{ (a }\OperatorTok{\textless{}}\StringTok{ }\DecValTok{1} \OperatorTok{+}\StringTok{ }\NormalTok{eps }\OperatorTok{\&}\StringTok{ }\NormalTok{b }\OperatorTok{\textless{}}\StringTok{ }\DecValTok{1} \OperatorTok{+}\StringTok{ }\NormalTok{eps) }\CommentTok{\# Degenerate case}
      \KeywordTok{return}\NormalTok{(}\KeywordTok{c}\NormalTok{(}\OtherTok{NA}\NormalTok{, }\OtherTok{NA}\NormalTok{))}
    \ControlFlowTok{if}\NormalTok{ (a }\OperatorTok{\textless{}}\StringTok{ }\DecValTok{1} \OperatorTok{+}\StringTok{ }\NormalTok{eps }\OperatorTok{\&}\StringTok{ }\NormalTok{b }\OperatorTok{\textgreater{}}\StringTok{ }\DecValTok{1}\NormalTok{) }\CommentTok{\# Left border case}
      \KeywordTok{return}\NormalTok{(}\KeywordTok{c}\NormalTok{(}\DecValTok{0}\NormalTok{, width))}
    \ControlFlowTok{if}\NormalTok{ (a }\OperatorTok{\textgreater{}}\StringTok{ }\DecValTok{1} \OperatorTok{\&}\StringTok{ }\NormalTok{b }\OperatorTok{\textless{}}\StringTok{ }\DecValTok{1} \OperatorTok{+}\StringTok{ }\NormalTok{eps) }\CommentTok{\# Right border case}
      \KeywordTok{return}\NormalTok{(}\KeywordTok{c}\NormalTok{(}\DecValTok{1} \OperatorTok{{-}}\StringTok{ }\NormalTok{width, }\DecValTok{1}\NormalTok{))}
    \ControlFlowTok{if}\NormalTok{ (width }\OperatorTok{\textgreater{}}\StringTok{ }\DecValTok{1} \OperatorTok{{-}}\StringTok{ }\NormalTok{eps)}
      \KeywordTok{return}\NormalTok{(}\KeywordTok{c}\NormalTok{(}\DecValTok{0}\NormalTok{, }\DecValTok{1}\NormalTok{))}
    
    \CommentTok{\# Middle case}
\NormalTok{    mode \textless{}{-}}\StringTok{ }\NormalTok{(a }\OperatorTok{{-}}\StringTok{ }\DecValTok{1}\NormalTok{) }\OperatorTok{/}\StringTok{ }\NormalTok{(a }\OperatorTok{+}\StringTok{ }\NormalTok{b }\OperatorTok{{-}}\StringTok{ }\DecValTok{2}\NormalTok{)}
\NormalTok{    pdf \textless{}{-}}\StringTok{ }\ControlFlowTok{function}\NormalTok{(x) }\KeywordTok{dbeta}\NormalTok{(x, a, b)}
    
\NormalTok{    l \textless{}{-}}\StringTok{ }\KeywordTok{uniroot}\NormalTok{(}
      \DataTypeTok{f =} \ControlFlowTok{function}\NormalTok{(x) }\KeywordTok{pdf}\NormalTok{(x) }\OperatorTok{{-}}\StringTok{ }\KeywordTok{pdf}\NormalTok{(x }\OperatorTok{+}\StringTok{ }\NormalTok{width),}
      \DataTypeTok{lower =} \KeywordTok{max}\NormalTok{(}\DecValTok{0}\NormalTok{, mode }\OperatorTok{{-}}\StringTok{ }\NormalTok{width),}
      \DataTypeTok{upper =} \KeywordTok{min}\NormalTok{(mode, }\DecValTok{1} \OperatorTok{{-}}\StringTok{ }\NormalTok{width),}
      \DataTypeTok{tol =} \FloatTok{1e{-}9}
\NormalTok{    )}\OperatorTok{$}\NormalTok{root}
\NormalTok{    r \textless{}{-}}\StringTok{ }\NormalTok{l }\OperatorTok{+}\StringTok{ }\NormalTok{width}
    \KeywordTok{return}\NormalTok{(}\KeywordTok{c}\NormalTok{(l, r))}
\NormalTok{  \}}

\NormalTok{  nw \textless{}{-}}\StringTok{ }\KeywordTok{kish\_ess}\NormalTok{(w)}
\NormalTok{  a \textless{}{-}}\StringTok{ }\NormalTok{(nw }\OperatorTok{+}\StringTok{ }\DecValTok{1}\NormalTok{) }\OperatorTok{*}\StringTok{ }\NormalTok{p}
\NormalTok{  b \textless{}{-}}\StringTok{ }\NormalTok{(nw }\OperatorTok{+}\StringTok{ }\DecValTok{1}\NormalTok{) }\OperatorTok{*}\StringTok{ }\NormalTok{(}\DecValTok{1} \OperatorTok{{-}}\StringTok{ }\NormalTok{p)}
\NormalTok{  hdi \textless{}{-}}\StringTok{ }\KeywordTok{getBetaHdi}\NormalTok{(a, b, width)}
\NormalTok{  hdiCdf \textless{}{-}}\StringTok{ }\KeywordTok{pbeta}\NormalTok{(hdi, a, b)}
\NormalTok{  cdf \textless{}{-}}\StringTok{ }\ControlFlowTok{function}\NormalTok{(n, p, t) \{}
    \ControlFlowTok{if}\NormalTok{ (p }\OperatorTok{==}\StringTok{ }\DecValTok{0} \OperatorTok{||}\StringTok{ }\NormalTok{p }\OperatorTok{==}\StringTok{ }\DecValTok{1}\NormalTok{)}
      \KeywordTok{return}\NormalTok{(}\KeywordTok{rep}\NormalTok{(}\OtherTok{NA}\NormalTok{, }\KeywordTok{length}\NormalTok{(t)))}
\NormalTok{    t[t }\OperatorTok{\textless{}=}\StringTok{ }\NormalTok{hdi[}\DecValTok{1}\NormalTok{]] \textless{}{-}}\StringTok{ }\NormalTok{hdi[}\DecValTok{1}\NormalTok{]}
\NormalTok{    t[t }\OperatorTok{\textgreater{}=}\StringTok{ }\NormalTok{hdi[}\DecValTok{2}\NormalTok{]] \textless{}{-}}\StringTok{ }\NormalTok{hdi[}\DecValTok{2}\NormalTok{]}
\NormalTok{    (}\KeywordTok{pbeta}\NormalTok{(t, a, b) }\OperatorTok{{-}}\StringTok{ }\NormalTok{hdiCdf[}\DecValTok{1}\NormalTok{]) }\OperatorTok{/}\StringTok{ }\NormalTok{(hdiCdf[}\DecValTok{2}\NormalTok{] }\OperatorTok{{-}}\StringTok{ }\NormalTok{hdiCdf[}\DecValTok{1}\NormalTok{])}
\NormalTok{  \}}
  \KeywordTok{wquantile\_generic}\NormalTok{(x, w, p, cdf)}
\NormalTok{\})}

\CommentTok{\# Weighted traditional quantile estimator}
\NormalTok{wquantile \textless{}{-}}\StringTok{ }\ControlFlowTok{function}\NormalTok{(x, w, probs, }\DataTypeTok{type =} \DecValTok{7}\NormalTok{) \{}
  \ControlFlowTok{if}\NormalTok{ (}\OperatorTok{!}\NormalTok{(type }\OperatorTok{\%in\%}\StringTok{ }\DecValTok{4}\OperatorTok{:}\DecValTok{9}\NormalTok{)) \{}
    \KeywordTok{stop}\NormalTok{(}\KeywordTok{paste}\NormalTok{(}\StringTok{"Unsupported type:"}\NormalTok{, type))}
\NormalTok{  \}}
\NormalTok{  cdf \textless{}{-}}\StringTok{ }\ControlFlowTok{function}\NormalTok{(n, p, t) \{}
\NormalTok{    h \textless{}{-}}\StringTok{ }\ControlFlowTok{switch}\NormalTok{(type }\OperatorTok{{-}}\StringTok{ }\DecValTok{3}\NormalTok{,}
\NormalTok{      n }\OperatorTok{*}\StringTok{ }\NormalTok{p,                   }\CommentTok{\# Type 4}
\NormalTok{      n }\OperatorTok{*}\StringTok{ }\NormalTok{p }\OperatorTok{+}\StringTok{ }\FloatTok{0.5}\NormalTok{,             }\CommentTok{\# Type 5}
\NormalTok{      (n }\OperatorTok{+}\StringTok{ }\DecValTok{1}\NormalTok{) }\OperatorTok{*}\StringTok{ }\NormalTok{p,             }\CommentTok{\# Type 6}
\NormalTok{      (n }\OperatorTok{{-}}\StringTok{ }\DecValTok{1}\NormalTok{) }\OperatorTok{*}\StringTok{ }\NormalTok{p }\OperatorTok{+}\StringTok{ }\DecValTok{1}\NormalTok{,         }\CommentTok{\# Type 7}
\NormalTok{      (n }\OperatorTok{+}\StringTok{ }\DecValTok{1} \OperatorTok{/}\StringTok{ }\DecValTok{3}\NormalTok{) }\OperatorTok{*}\StringTok{ }\NormalTok{p }\OperatorTok{+}\StringTok{ }\DecValTok{1} \OperatorTok{/}\StringTok{ }\DecValTok{3}\NormalTok{, }\CommentTok{\# Type 8}
\NormalTok{      (n }\OperatorTok{+}\StringTok{ }\DecValTok{1} \OperatorTok{/}\StringTok{ }\DecValTok{4}\NormalTok{) }\OperatorTok{*}\StringTok{ }\NormalTok{p }\OperatorTok{+}\StringTok{ }\DecValTok{3} \OperatorTok{/}\StringTok{ }\DecValTok{8}  \CommentTok{\# Type 9}
\NormalTok{    )}
\NormalTok{    h \textless{}{-}}\StringTok{ }\KeywordTok{max}\NormalTok{(}\KeywordTok{min}\NormalTok{(h, n), }\DecValTok{1}\NormalTok{)}
    \KeywordTok{pmax}\NormalTok{(}\DecValTok{0}\NormalTok{, }\KeywordTok{pmin}\NormalTok{(}\DecValTok{1}\NormalTok{, t }\OperatorTok{*}\StringTok{ }\NormalTok{n }\OperatorTok{{-}}\StringTok{ }\NormalTok{h }\OperatorTok{+}\StringTok{ }\DecValTok{1}\NormalTok{))}
\NormalTok{  \}}
  \KeywordTok{wquantile\_generic}\NormalTok{(x, w, probs, cdf)}
\NormalTok{\}}
\end{Highlighting}
\end{Shaded}

\newpage

\printbibliography

\end{document}